%% file: Main.tex
\documentclass[final,5p,times,twocolumn]{elsarticle}

\usepackage{amssymb}
\usepackage{amsmath}

\usepackage{stfloats}

\usepackage{xcolor}
\usepackage{soul}
\usepackage[normalem]{ulem}  

\usepackage{geometry}
\usepackage{etoolbox}
\patchcmd{\MaketitleBox}
  {\footnotesize\itshape\elsaddress\par\vskip36pt}
  {\footnotesize\itshape\elsaddress\par
   \vskip2pt
   \normalsize\upshape
   Project page: \href{https://wy-blacksheep.github.io/projects/physical-surface/}{\textcolor[HTML]{455fd4}{https://wy-blacksheep.github.io/projects/physical-surface/}}\par
   \vskip18pt}
  {}{}

\usepackage[hidelinks]{hyperref}
\makeatletter
\providecommand{\doi}[1]{%
  \begingroup
    \let\bibinfo\@secondoftwo
    \urlstyle{rm}%
    \href{https://doi.org/#1}{%
      doi:\discretionary{}{}{}%
      \nolinkurl{#1}%
    }%
  \endgroup
}
\makeatother

\journal{International Journal of Human-Computer Studies}

\begin{document}

\begin{frontmatter}



\title{Physical Surfaces Make Touch Interactions in Virtual Reality\\ Precise, Efficient, and Bimanual} 


\author{Wen Ying\corref{cor1}}
\ead{wenying@virginia.edu}
\ead[url]{https://wy-blacksheep.github.io/}

\author{Seongkook Heo}
\ead{seongkook@virginia.edu}
\ead[url]{https://seongkookheo.com/}

\cortext[cor1]{Corresponding author.}

\affiliation{organization={University of Virginia},
            city={Charlottesville},
            postcode={22903}, 
            state={Virginia},
            country={USA}}

\begin{abstract}

Virtual reality (VR) systems can enable convenient hand-based interactions across diverse work scenarios. 
However, mid-air gestures lack tactile feedback and a physical reference surface to support the hand. This absence of haptic grounding can cause significant challenges in achieving precise and efficient touch interactions. This paper investigates the effect of different types of hand-grounded haptic feedback on the touch performance of VR tasks that demand high precision, such as selecting, tracing, and sketching. We compared three levels of haptic feedback: 1) No Haptic Feedback, where only visual feedback was provided; 2) Tactile Feedback, where users received vibrotactile and pressure feedback upon touching a virtual surface; 3) Physical Surface, where users interacted with a portable and tangible surface. 

Our study found that portable physical surfaces enabled the best selection precision, tracing efficiency, and sketch quality. Furthermore, participants showed increased bimanual hand utilization when engaging with a physical surface during tasks. These observed behaviors corresponded to participants' preference for interacting with physical surfaces, attributed to a better sense of confidence and control.
\end{abstract}




\begin{keyword}
Touch interaction \sep Portable physical surfaces \sep Haptic feedback \sep Virtual reality



\end{keyword}

\end{frontmatter}


\input{Introduction}
\input{RelatedWork}
\input{Method}
\input{Results}
\input{Discussion}
\input{Conclusion}

\section{Acknowledgment}
This work has been supported by the UVA Research Innovation Award. 

\appendix
\input{Appendix}

\bibliographystyle{elsarticle-num-names} 
\bibliography{reference}

\end{document}

%% file: Introduction.tex
\section{INTRODUCTION}

With the development of portable consumer virtual reality (VR) devices, such as Apple Vision Pro and Meta Quest 3, VR is now used for work and entertainment in diverse environments ranging from home and office settings \cite{10.1145/3313831.3376724} to public spaces such as restaurants \cite{10.1145/3365610.3365647} and transportation \cite{10.1145/3349263.3351330, ofek2020towards}.

As VR devices allow users to utilize the expansive virtual space in diverse environments, efforts have been made to enhance the potential of VR for professional work by integrating high-performance input devices like keyboards and mice. However, the lack of a surface to place these devices or the burden of carrying them can limit their usability \cite{grubert2018office}. 
Without such input devices, users often rely on mid-air hand interactions. Both commercial VR headsets and academic research have popularized the use of bare hands for frequent surface-based interactions in VR, such as menu interaction and window manipulation \cite{biener2020breaking, Lee2018_window_fingertip, 10.1145/3461778.3462076}. While this approach offers benefits such as availability, intuitiveness, convenience, and versatility \cite{xiao18mrtouch}, the lack of haptic feedback in mid-air gestures can pose challenges for precision-dependent interactions \cite{8943750, NormalTouch_and_TextureTouch, evaluation_sketching_Arora}.

Researchers have developed various haptic feedback techniques to improve mid-air hand interactions and investigate their effects on interaction performance. One of the popular approaches is to provide tactile feedback to a user’s finger so that the user can feel the contact with the virtual surface via vibrotactile \cite{5951857, 6758358} and pressure feedback \cite{8798255, 10.1145/2929464.2929474, Ma2015DesignAO, 10.1145/3290605.3300301}. Additionally, studies have shown that combining tactile and pressure feedback and using them further improves the accuracy and speed of selecting, tapping, and tracing \cite{HapThimble, NormalTouch_and_TextureTouch}. 
Another well-investigated method to enhance hand-based VR interactions is using a physical surface. Having a grounded surface, such as furniture surfaces or fixed touchscreens, has been found to benefit hand-based touch interaction \cite{10.1145/332040.332494, xiao18mrtouch, 8943750} and pen-based sketching \cite{evaluation_sketching_Arora, investigating_sketching_Israel} by improving the manipulation precision and speed. 

\begin{figure}[!tb]
\centering
\includegraphics[width=\linewidth]{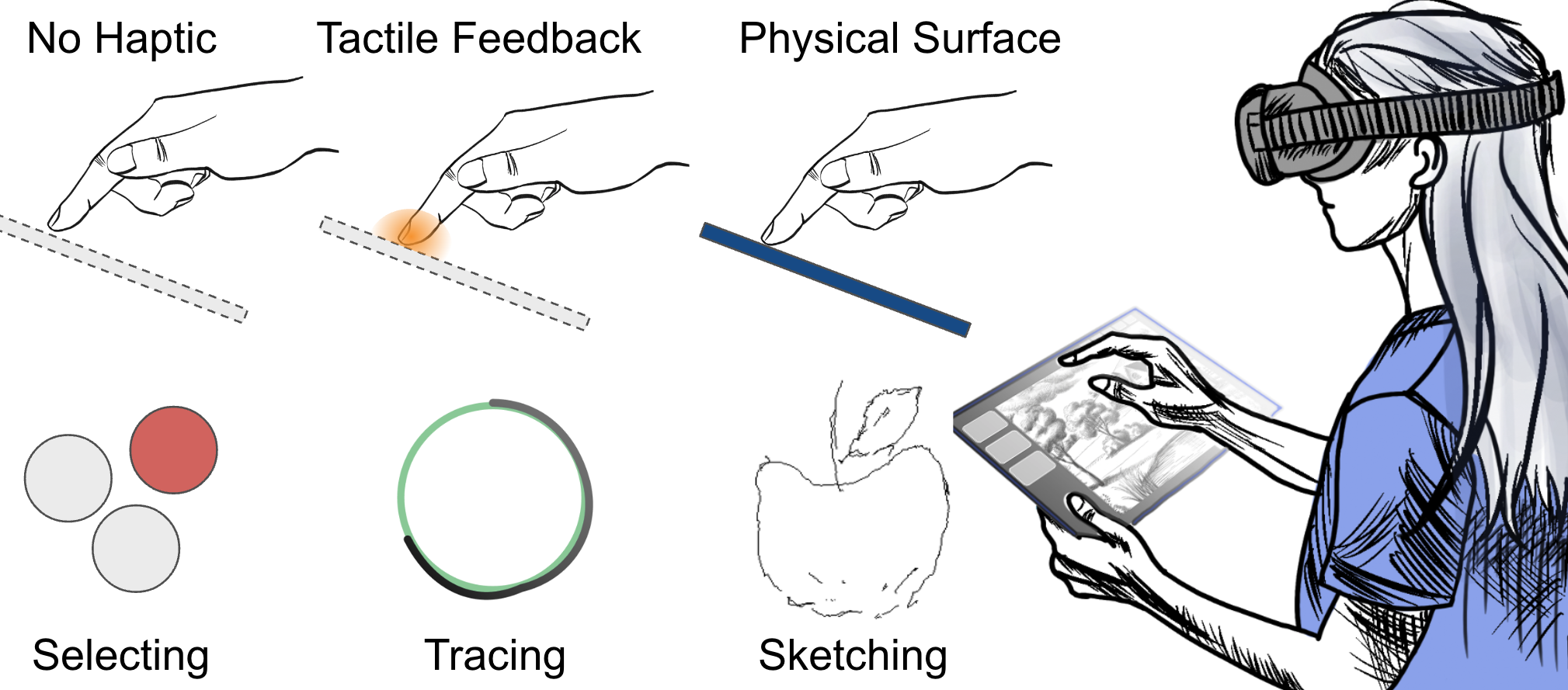}
\caption{This study investigates the effects of visual feedback, tactile feedback, and physical surfaces on precise touch interactions for selecting, tracing, and sketching in VR.}
\vspace{-5pt}
\label{physical_vrsketch_teaser}
\end{figure}

Although previous studies have highlighted the performance benefits of using grounded physical surfaces for touch interactions, large and fixed surfaces may not always be available for users. Portable surfaces, such as smartphones and tablets, provide a flexible alternative. Some 3D design systems have incorporated digital drawing tablets in VR to enable advanced model manipulation and detailed sketching \cite{VRSketchIn, SymbiosisSketch, TabletInVR}. However, findings from studies on grounded surfaces may not directly apply to portable surfaces due to their 6-DoF mobility and the involvement of both hands in holding the device. Additionally, while pens are commonly used for precise sketching and other complex interactions on tablets, their usability can be limited by their availability and the lack of multi-touch functionality. By exploiting the availability and versatility of human hands \cite{xiao18mrtouch, human_hand_function}, we aim to investigate the potential of hand interactions for tasks requiring precision in VR. Our research focuses on how different haptic modalities affect hand-based touch interactions on portable surfaces in VR, their impact on the performance of tasks across varying complexity, and their influence on user behavior.

This paper shares insights from a user study that investigated the effects of physical surfaces on precise touch interactions in VR by analyzing precision, speed, and hand behavior during selecting, tracing, and sketching tasks (Figure \ref{physical_vrsketch_teaser}). We compared these results with a baseline condition without any haptic feedback and tactile feedback provided by pressure and vibrotactile actuators attached to a haptic glove. We found that physical surfaces benefit participants most in all tasks, while tactile feedback outperforms the baseline condition without haptic feedback. Compared with interacting with tactile feedback, using physical surfaces improved selection accuracy by 51.2\%; tracing precision by 20.3\%, tracing continuity by 86.8\%, and tracing speed by 18.3\%; stroke smoothness by 21.6\%, stroke continuity by 21.2\%, and sketch clarity by 17.5\%. In addition, we found that participants' non-dominant hands moved the physical surface at a $\sim100\%$ larger distance to adjust the surface during tracing and sketching. The dominant hand reduced its total movements but hovered at a $\sim150\%$ higher position and approached the surface faster. These observed behaviors may suggest enhanced engagement and coordination between both hands to share the workload, aligning with participants' preference for interacting with physical surfaces, which they reported as providing greater confidence and control. 

Integrating physical surfaces in VR for touch interactions significantly enhances task precision and efficiency, potentially benefiting from the coordinated bimanual behaviors that physical surfaces facilitate. We envision the increased utilization of hand interactions to support more complex tasks involving subtle touching and precise moving and provide users with more intuitive, convenient, and professional experiences in their daily VR activities.

%% file: RelatedWork.tex
\section{RELATED WORK}
We review research investigating haptic feedback techniques to enhance immersive hand-based manipulation, the utilization of physical objects to facilitate precise interactions in VR, and the impact of bimanual actions on touch-based interaction.

\subsection{Haptic Devices for Immersive Hand Manipulation}
Both commercial VR headsets, such as Apple Vision Pro and Meta Quest 3, and academic research have popularized the use of bare hands for frequent surface-based interactions in VR, including menu navigation and window manipulation \cite{biener2020breaking, Lee2018_window_fingertip, 10.1145/3461778.3462076}. While hands provide intuitive and convenient touch interaction in the current VR systems, the lack of tactility has been a significant challenge for improving interaction efficiency \cite{bouzbib2021can, 10.1007/978-3-642-40483-2_3}. This problem has prompted the development of various haptic feedback methods designed to simulate touch sensations through tactile and pressure feedback.

One of the most popular methods of simulating the sense of contact is using vibrotactile feedback. Previous studies have produced per-finger vibrations when the finger touches the virtual object \cite{5951857, 6758358, 10.1145/2614066.2614079}. Ultrasonic actuators have also been used to generate localized tactile feedback in the air and produce the feeling of contact \cite{10.1145/2501988.2502018, 10.1145/2821592.2821593, long2014rendering}. However, they cannot affect users' physical hand movements, such as preventing the finger from penetrating a virtual surface. 

To address this problem, body-grounded haptic devices have been leveraged to provide pressure feedback that mimics the resistance of the physical surface. These devices can physically push, pull, or stop the finger based on the virtual context using wires \cite{8798255, 10.1145/3313831.3376470}, mechanical systems \cite{HapThimble, strasnick2018haptic}, and electrostatic brakes \cite{teng2022prolonging, choi2016wolverine, 10.1145/3242587.3242657, senseglove}. 

In addition, research has found that combining tactile and pressure feedback can further enhance the interaction accuracy and speed \cite{NormalTouch_and_TextureTouch}. The HapThimble system \cite{HapThimble}, which simulated a touchable surface through tactile and pressure feedback, achieved state-of-the-art touch performance in terms of selection speed and accuracy. However, since the HapThimble is a rigid cylindrical device that wrapped the manipulation finger, it does not support natural finger bending and might affect touch performance \cite{wacharamanotham2014understanding}. Its 25 mm diameter is also too large for small-target interaction, exceeding the recommended selection UI sizes from Google and Apple (7–10 mm) and prior research (11.6–13 mm) \cite{pointing_in_vr, wang2009empirical}. The large cylindrical body can occlude small targets, making precise selection tasks challenging and less reliable.

\subsection{Using Physical Objects to Enhance Interaction Performance}
While numerous haptic feedback techniques have been developed to enhance immersive VR experiences, some research has focused on blending physical and virtual worlds to enrich user experience \cite{roo2017one}, others have aimed to improve interaction performance by integrating physical objects into VR. For instance, MRTouch \cite{xiao18mrtouch} enabled detecting touch input on a physical surface for precise interaction. Studies found that typing \cite{8943750} and manipulating virtual objects \cite{10.1145/332040.332494, 10.1145/2512349.2512805} can also benefit from using a table. Encountered-type haptic systems have been developed to increase accessibility by using robotic arms or mobile robots with physical surfaces. These systems dynamically present physical surfaces to align with virtual interactions \cite{horie2021encounteredlimbs, mortezapoor2023cobodeck, gomi2023ubisurface, weng2025hitaround}. Although encountered systems provide responsive haptic experiences, their focus is primarily on enhancing immersion and realism rather than improving task performance, and their reliance on complex, large-scale robotic setups limits practicality.

In contrast, holding a portable physical proxy in hand offers a flexible alternative, allowing users to interact with the corresponding virtual target from various positions and orientations. Research has also shown that, compared to grounded physical objects, in-hand physical objects enable more efficient interactions, such as selection and docking \cite{mine1997moving, lindeman1999hand}. Portable controllers, such as TORC \cite{10.1145/3290605.3300301}, NormalTouch and TextureTouch \cite{NormalTouch_and_TextureTouch}, provided users with small physical objects for finger touch or pressing, which enhanced performance of grabbing \cite{10.1145/3290605.3300301}, targeting and tracing \cite{NormalTouch_and_TextureTouch}. However, on-surface finger movements in these systems are either stationary or restricted to a tiny area. Such constraints can make fine-grained continuous inputs, such as writing or sketching, difficult because the finger lacks the traction and friction available on larger surfaces.

Research on more complex touch interactions in VR often uses larger surfaces, such as drawing on virtual 3D models. However, pen-based sketching in VR remains less precise than traditional 2D drawing \cite{investigating_sketching_Israel}. To mitigate this performance decrease, studies have introduced planar physical surfaces that improve stroke precision and smoothness by providing haptic feedback and preventing penetration into virtual space \cite{evaluation_sketching_Arora}. Systems such as VRSketchIn \cite{VRSketchIn}, TabletInVR \cite{TabletInVR}, and SymbiosisSketch \cite{SymbiosisSketch} leverage tablets or touchscreens to enhance detailed sketching. While such setups offer high precision, they require additional hardware and integration with VR systems. In contrast, hand-based interaction is more accessible and device-free, supporting ad-hoc and multi-touch input for dexterous manipulation. With advances in hand tracking, it is worth exploring whether users could leverage everyday physical surfaces in their environment to achieve high-performance hand-based touch interactions such as sketching. 

However, the effect of physical surfaces on finger-based VR sketching remains unclear, as results from pen-based interactions on grounded surfaces \cite{evaluation_sketching_Arora} may not directly apply to hand interactions on portable surfaces. Finger-based input differs from pen-based input in resolution, contact point, and stroke size \cite{finger_pen_stroke, 10.1145/2797138}, since using a tool fundamentally changes task execution \cite{34763}. Additionally, surface mobility can further influence touch performance \cite{mine1997moving, lindeman1999hand}.

\subsection{Benefits of Asymmetric Bimanual Interactions}
When manipulating a portable surface, users typically coordinate both hands in distinct roles, where the non-dominant hand holds the object for the dominant hand to interact with it. Guiard’s Kinematic Chain model \cite{guiard1987asymmetric} explains that the non-dominant hand (NDH) establishes a frame of reference, providing context and stability for the dominant hand (DH) to perform precise actions. This coordination leverages the benefits from human proprioception, as users can more easily perceive the relative position of their hands than the individual position of each hand \cite{10.1145/320719.322599, mine1997moving, hinckley1998two}. Hinckley further emphasized the role of the NDH in guiding the DH, supporting the concept that asymmetric bimanual actions are naturally efficient and intuitive \cite{10.1145/258549.258571}. 

Such asymmetric bimanual interactions are essential for enhancing task performance, particularly in terms of precision and speed \cite{10.1145/320719.322599}. For example, a previous study showed that using both hands to interact with a physical prop helped people effectively manipulate a neurosurgical model \cite{10.1145/191666.191821}. Complex and compound tasks, such as drawing and color selection tasks, can also benefit from bimanual interactions, leading to improved precision and reduced task completion time \cite{10.1145/191666.191808, 10.1145/320719.322599}. While these studies focused on real-world interactions, this insight into bimanual interactions can also be applied to VR. When both hands work together, they provide richer sensory cues than unimanual control, enhancing the user's spatial awareness and improving object manipulation and hand repositioning in VR \cite{Hinckley1997FrameReference}. This suggests that using the NDH to hold a virtual object while the DH acts on it can improve on-surface touch interactions in VR.

\subsection{Discussion and Summary}
Research has explored various haptic feedback techniques to enhance user experience and task performance in VR. Prior work has largely focused on fingertip-level feedback or localized tactile effects to create realistic sensations of contact and texture \cite{5951857, 10.1145/2501988.2502018, 8798255, teng2022prolonging, HapThimble}. Systems such as NormalTouch and TextureTouch \cite{NormalTouch_and_TextureTouch} and HapThimble \cite{HapThimble} render tactile feedback through small active surfaces attached to the fingertip, enabling users to perceive surface contact or local shape. However, these devices constrain finger motion to a limited area and rely primarily on wrist and arm movement. While some systems, such as MRTouch \cite{xiao18mrtouch} can support surface touch in a more realistic way, grounded environments are less flexible and may not always be available.

In contrast, our study extends this body of work by systematically investigating how portable physical surfaces can support tasks requiring different levels of precision. Unlike prior systems that isolate feedback to the fingertip or depend on stationary proxies, our lightweight, portable surface leverages bimanual proprioception, allowing the non-dominant hand to anchor and stabilize the interaction space while the dominant hand performs fine-grained gestures on the surface. This configuration enables continuous, tablet-like interaction that remains mobile and device-free.

Furthermore, our work systematically compares usability and performance outcomes across visual-only, tactile, and physical feedback conditions, revealing how different haptic modalities influence user performance and experience across tasks. While previous studies often evaluated a single haptic mode or task type \cite{10.1145/332040.332494, HapThimble, NormalTouch_and_TextureTouch, xiao18mrtouch, 10.1145/3290605.3300301}, our design spans multiple interaction contexts, providing a more holistic understanding of how physical surfaces impact bimanual interaction performance in VR.

Overall, this study contributes to a broader perspective on bimanual, surface-based interaction by demonstrating that a simple hand-held physical surface can improve both user performance and user experience across multiple input tasks.

%% file: Method.tex
\section{COMPARATIVE HAPTIC FEEDBACK STUDY}

\begin{figure*}[t]
\centering
\includegraphics[width=0.8\textwidth]{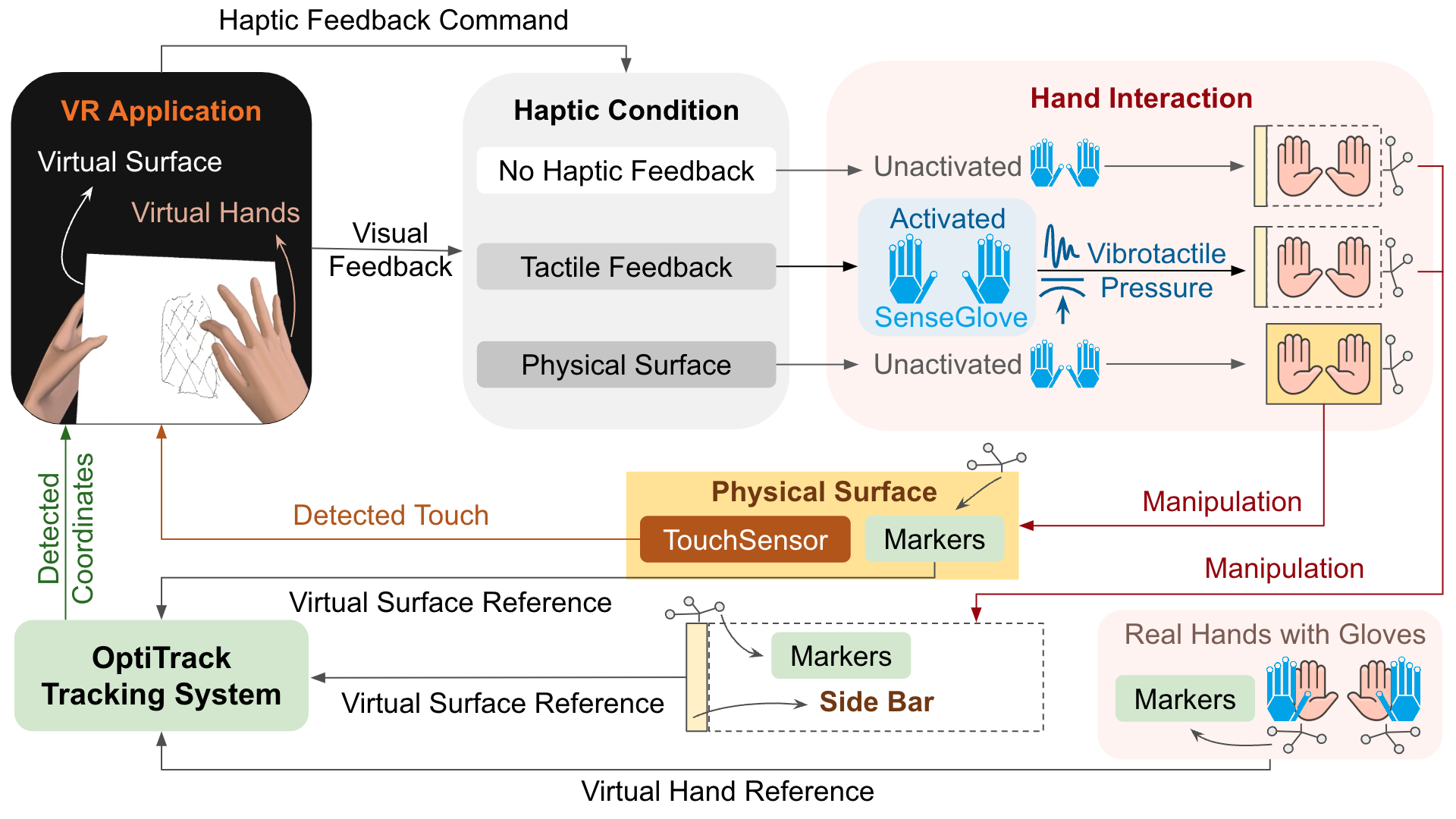}
\vspace{-6pt}
\caption{System Design Overview.}
\label{system}
\end{figure*}

\begin{figure*}[t]
\centering
\includegraphics[width=0.75\textwidth]{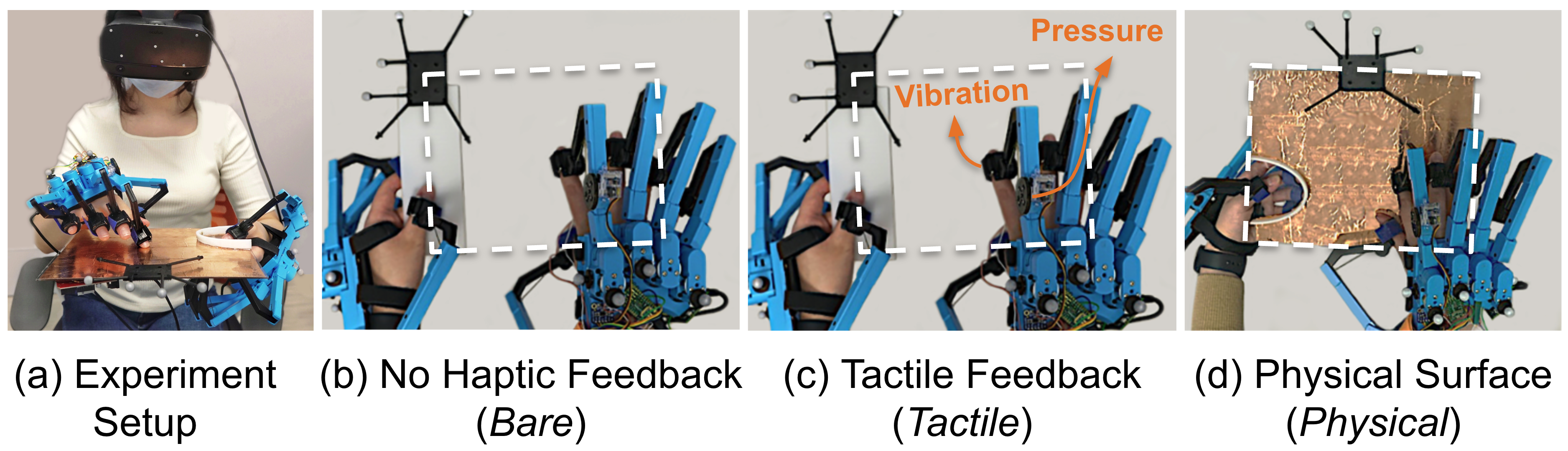}
\caption{The experimental setup (a) and three haptic feedback conditions: (b) no haptic feedback (Bare); (c) tactile feedback (Tactile); (d) physical surface (Physical). The white dashed line represents the virtual surface presented in VR.}
\vspace{-3pt}
\label{haptic_feedback}
\end{figure*}

This study investigated how much benefit an in-hand physical surface can bring to touch interaction in precision-required VR tasks. We developed a system to evaluate the effects of three types of haptic feedback (Figure \ref{system}): \textit{no haptic feedback (Bare)}, \textit{tactile feedback (Tactile)}, and \textit{physical surface (Physical)}, on task performance and user behavior. Three tasks with different manipulation complexities were tested: selecting, tracing, and sketching.

\subsection{Hypotheses}
\textit{No haptic feedback} mimicked the mid-air, free-hand interaction that is common in commercial VR devices (e.g., Apple Vision Pro, Meta Quest 3), where only visual feedback is provided. \textit{Tactile feedback} synthetically generated vibrotactile and pressure feedback, inspired by prior work that achieved the state-of-the-art clicking performance in VR \cite{HapThimble}. \textit{Physical surface} supported touch interactions that replicated how people manipulate physical objects in the real world.

Based on the characteristics of the three haptic conditions, this study aimed to evaluate the following two hypotheses:
\begin{itemize}
	\item \textit{Physical surface} will most effectively enhance task performance in VR, followed by \textit{tactile feedback}, with \textit{no haptic feedback} as the baseline.
	
	\item User hand behaviors will vary significantly depending on whether they are interacting with a \textit{physical surface}, \textit{tactile feedback}, or \textit{no haptic feedback}.
\end{itemize}

\subsection{Study Setup}
The design of the system utilized in this study is illustrated in Figure \ref{system}, and the experimental setup is shown in Figure \ref{haptic_feedback}. With different haptic feedback, this study investigated their effects on precise interactions that leverage users' free finger movements.

\subsubsection{Tracking}
To precisely track finger articulations and positions, we used a pair of commercial hand-tracking gloves (i.e., SenseGlove DK1 exoskeleton) as demonstrated in PalmEX \cite{bouzbib2023palmex}. Additionally, as shown in Figure \ref{system}, the OptiTrack system was employed to track the overall apparatus, the handheld surface, and the interaction dynamics, such as identifying where the interacting finger made contact with the surface. 

\begin{figure}[!tb]
\centering
\includegraphics[width=\linewidth]{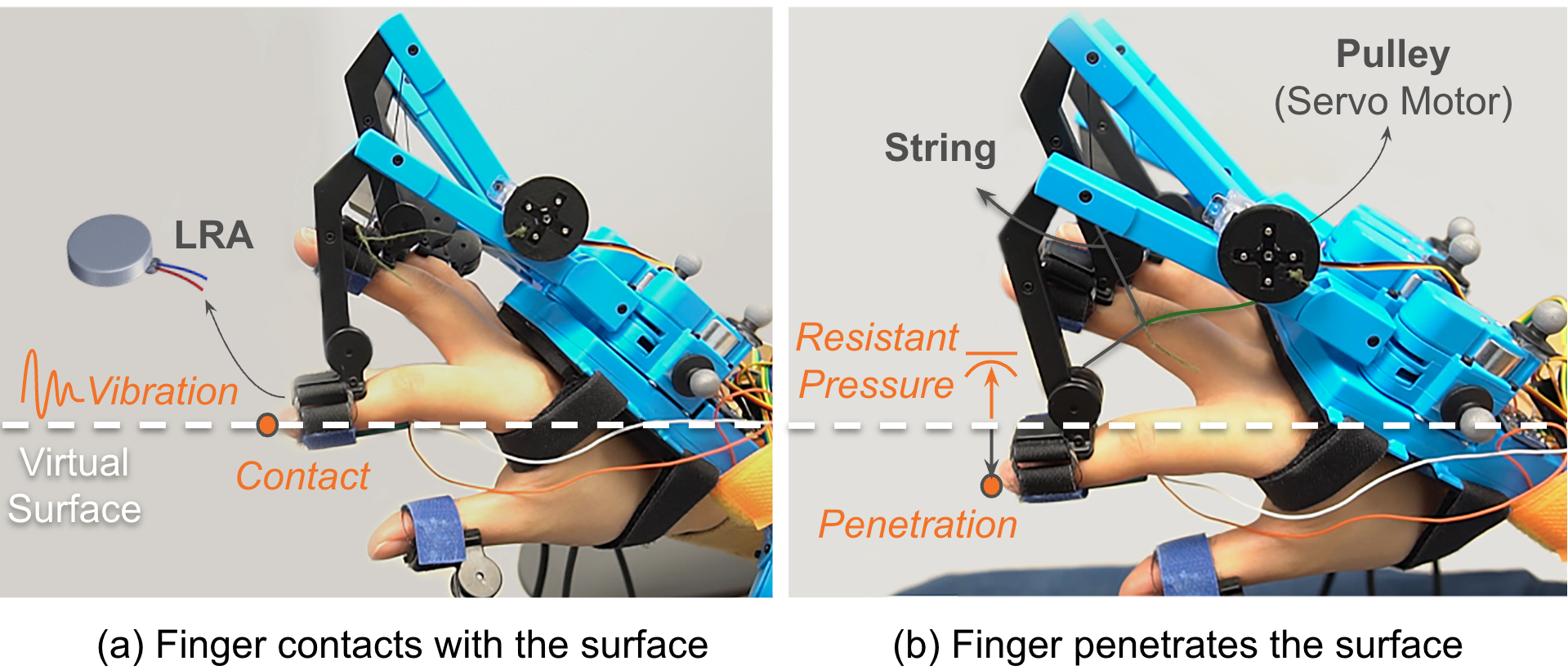}
\caption{Virtual haptic feedback activation: (a) The LRA generates contact vibrations to simulate tactile feedback when the finger touches the virtual surface; (b) the pulley rotated by a servo motor pulls the string to generate pressure when the finger penetrates the virtual surface. The white dashed line shows the virtual surface in the horizontal view.}
\vspace{-3pt}
\label{tactile_feedback}
\end{figure}

\begin{figure*}[!tb]
\centering
\includegraphics[width=0.7\textwidth]{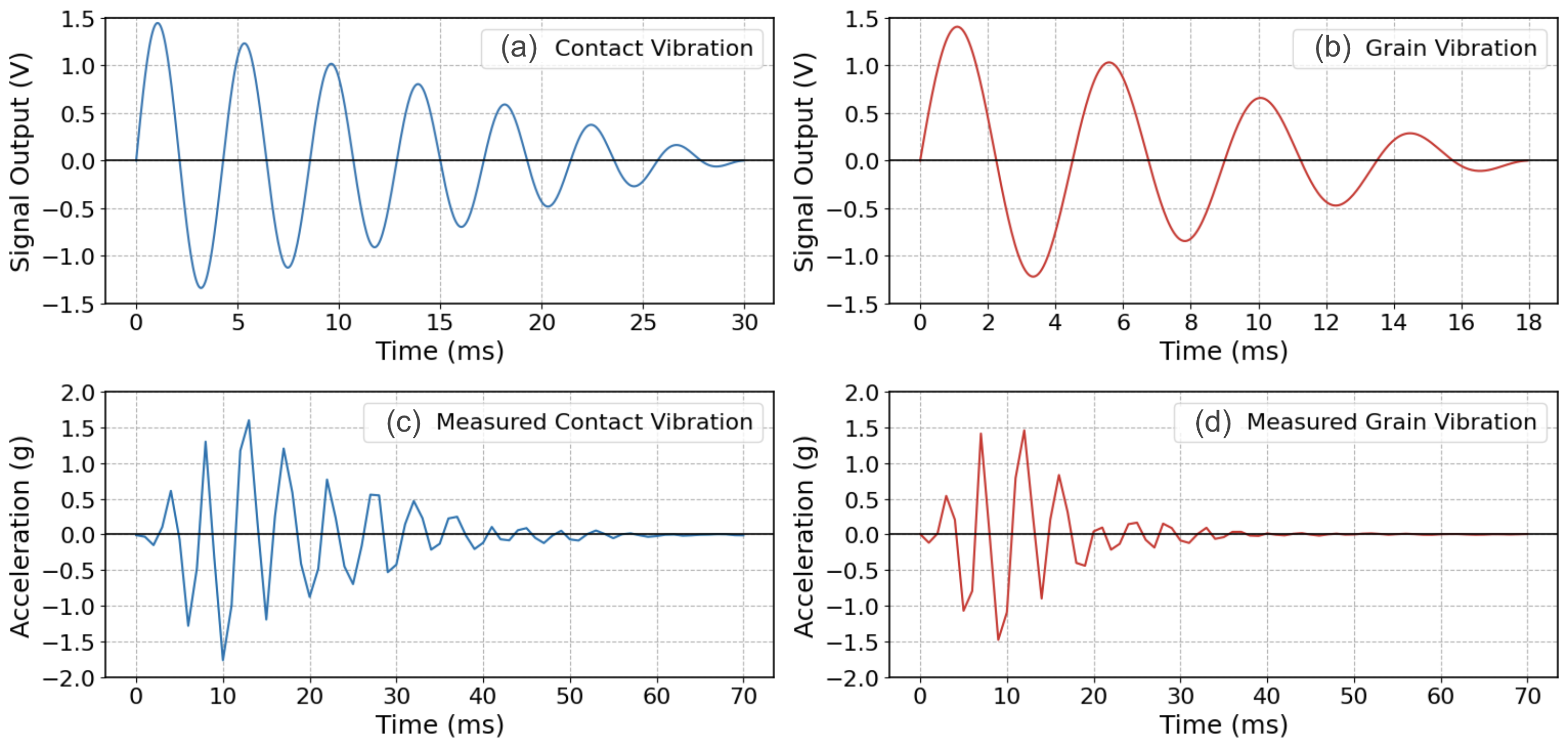}
\caption{Control signals to generate (a) contact and (b) grain vibrations, and the corresponding acceleration measurements for (c) contact and (d) grain vibrations.}
\vspace{-5pt}
\label{tactile_vibration}
\end{figure*}

\subsubsection{Apparatus}
Participants wore an Oculus Rift S VR headset ($\sim 500$ g) and a pair of SenseGlove DK1 devices (316 g per glove; Figure \ref{haptic_feedback}a) while seated comfortably in a chair. Although no table was provided, participants could rest their arms on the chair’s armrests. This setup reduced the need to extend the arms forward continuously, thereby alleviating the “gorilla arm” effect \cite{hansberger2017dispelling}. To ensure fairness across conditions, participants were instructed not to place the physical plate on their laps, so that the physical surface condition would still reflect a ``hand-grounded'' setup comparable to the other haptic conditions. The gloves and handheld surfaces had reflective markers on them and were tracked by OptiTrack trackers at 360 Hz, and the mean tracking error was 0.28 mm. We used the hand model provided by the SenseGlove SDK, which also captures finger phalanx positions and articulations at 400 Hz using inverse kinematics (IK). Since participants interacted with the index finger of their dominant hand, the interaction point was defined as the center of the distal phalanx tip of the index finger. This point was then mapped onto the corresponding virtual surface to determine the selection or stroke center. Virtual hands and a $255 \times 205$ mm virtual surface were rendered at $\sim50$ Hz to provide participants with visual feedback throughout the experiment (Figure \ref{system}). The experiment software was developed using Unity and ran on a laptop with an Intel Core i7 processor and an NVIDIA GTX 1060 GPU.

We required participants to wear the SenseGlove across all experimental conditions to maintain consistency in tracking and data collection. In the Bare and Physical conditions, the SenseGlove was used solely to track hand gestures without providing haptic feedback. In contrast, during the Tactile condition, the SenseGlove was activated to deliver tactile feedback while supporting hand tracking.

\subsubsection{No Haptic Feedback (Bare)}
In the \textit{Bare} condition, participants received only visual feedback, with no haptic feedback provided. As shown in Figure \ref{haptic_feedback}b, participants held a narrow bar with their non-dominant (left) hand. This bar served as a reference for situating a virtual handheld surface in the environment. Participants used their dominant (right) index finger to perform tasks on the virtual surface. 

Touch detection in this condition was achieved by monitoring collisions between the interacting fingertip and the virtual surface. The virtual finger representation was rendered to penetrate the surface, as no physical constraint was applied.

Given our study's primary focus on evaluating precise interaction performance, we presented participants with unaltered hand actions without visual manipulation. This consistent, unmodified visual feedback was maintained across all conditions.

\subsubsection{Tactile Feedback (Tactile)}

We adopted the same tactile feedback (vibrotactile + pressure) developed in HapThimble \cite{HapThimble}. However, the HapThimble is designed to be worn on the entire finger, which limits finger bending and is too bulky for small targets. We aimed to replicate HapThimble's tactile and pressure feedback while enabling users to bend their fingers freely for precise and dexterous interactions \cite{wacharamanotham2014understanding}.

Inspired by PalmEx \cite{bouzbib2023palmex}, which modified haptic gloves to support flexible finger manipulation, we adapted SenseGlove \cite{senseglove} to replicate HapThimble’s feedback by integrating several actuators controlled through a Teensy 4.0 microcontroller (Figure \ref{haptic_feedback}c). To deliver vibrotactile feedback directly to the fingertip, a Linear Resonant Actuator (LRA; Jinlong Machinery, G0832012; $\sim10$ ms latency, with a Texas Instruments DRV2605L motor driver) was attached to the fingertip pad via Velcro tape, allowing participants to feel contact sensations under the fingertip when touching or moving on the surface (Figure \ref{tactile_feedback}a). To render the feeling of initial finger contact (collision), we used the same contact vibration pattern as HapThimble \cite{HapThimble}: a 235 Hz sine wave that decays to 0 over 30 ms (Figure \ref{tactile_vibration}a). If the finger kept moving on the surface, we provided continuous contact feedback through a shorter vibration pattern that decayed to 0 over 18 ms (Figure \ref{tactile_vibration}b), similar to the grain vibration effect in HapThimble \cite{HapThimble}. Figure \ref{tactile_vibration} also shows accelerations for (c) contact and (d) grain vibrations, measured with an InvenSense ICM-20948 IMU at a 1000 Hz sampling rate.

Additionally, a pulley rotated by a servo motor (Hitec, HS-40; $\sim60$ ms latency) was integrated with the SenseGlove to simulate variable pressure proportional to penetration depth, allowing participants to perceive the depth of penetration and adjust their finger movements accordingly (Figure \ref{tactile_feedback}b). When the finger penetrated the virtual surface, the pulley simulated surface resistance by applying pressure through a string connected to the SenseGlove finger joint. The pressure level was linearly mapped to the penetration depth, up to a maximum of 30 mm, approximately the length of a fingertip. Beyond 30 mm, the pressure was capped at 13 N, simulating a hard surface limit.

The handheld surface, the tracking mechanism, and the touch detection method were identical to those used in the \textit{Bare} condition, ensuring consistency across conditions.

\subsubsection{A Physical Surface (Physical)}
In the \textit{Physical} condition, participants held a $255 \times 205$ mm transparent acrylic plate weighing 185 g in their left hand, as shown in Figure \ref{haptic_feedback}d. To enable touch detection on the physical surface \cite{HapThimble}, the acrylic plate was equipped with a capacitive touch sensor made from copper tape. To prevent false touches by the non-dominant hand, a portion of the plate was left uncovered by the touch sensor and surrounded by a 10 mm high barrier, as illustrated in Figure \ref{haptic_feedback}d. The location of the fingertip on the surface was consistently tracked using the OptiTrack system, as in the \textit{Bare} and \textit{Tactile} conditions, ensuring uniform interaction tracking across all conditions.

\subsection{Tasks}


The experiment performed three tasks with various manipulation complexities: selecting (Figure \ref{selection_task}), tracing (Figure \ref{tracing_task}), and sketching (Figure \ref{sketching_task}).

\subsubsection{Selecting}

\begin{figure}[t]
\centering
\includegraphics[width=\linewidth]{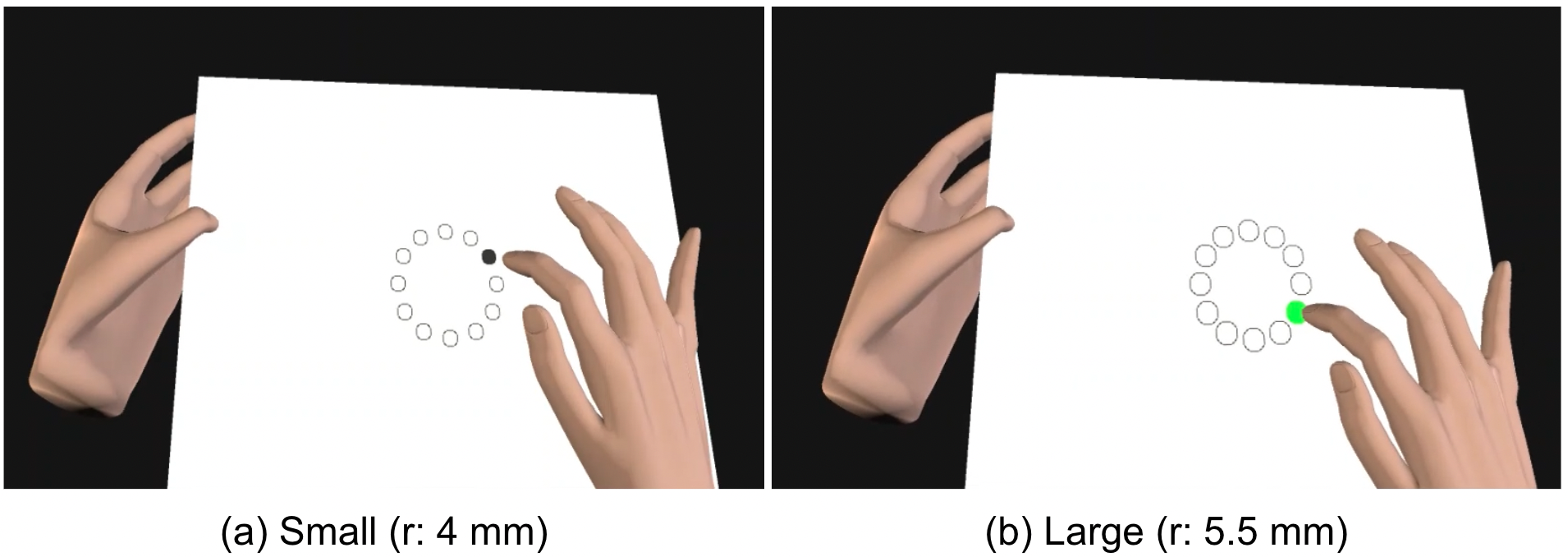}
\caption{Screenshots of the two selection tasks performed in this study: (a) small targets with radii of 4 mm and (b) large targets with radii of 5.5 mm.}
\vspace{-3pt}
\label{selection_task}
\end{figure}

Selection is one of the fundamental tasks for evaluating touch performance in VR \cite{HapThimble, xiao18mrtouch}. In this study, we implemented 2D Fitts' law selection tasks \cite{pointing_in_vr}, featuring small circular targets with radii of 4 mm (Figure \ref{selection_task}a) and large targets with radii of 5.5 mm (Figure \ref{selection_task}b). The distance between the sequential targets was fixed at 50 mm.

The chosen target sizes align with the minimum touch target sizes recommended by industry UI design guidelines from Google and Apple (3.5–5 mm) and academic research (5.8 mm) \cite{pointing_in_vr}.

As shown in Figure \ref{selection_task}a, the target initially appeared in black. Upon a successful selection, the target's color changed to green, as shown in Figure \ref{selection_task}b, and the subsequent target appeared in black. In the case of a selection error, the cursor under the interacting finger turned red, and the participant proceeded to the next target. Participants were instructed to complete the selections as quickly and accurately as possible.

\subsubsection{Tracing}

\begin{figure}[t]
\centering
\includegraphics[width=\linewidth]{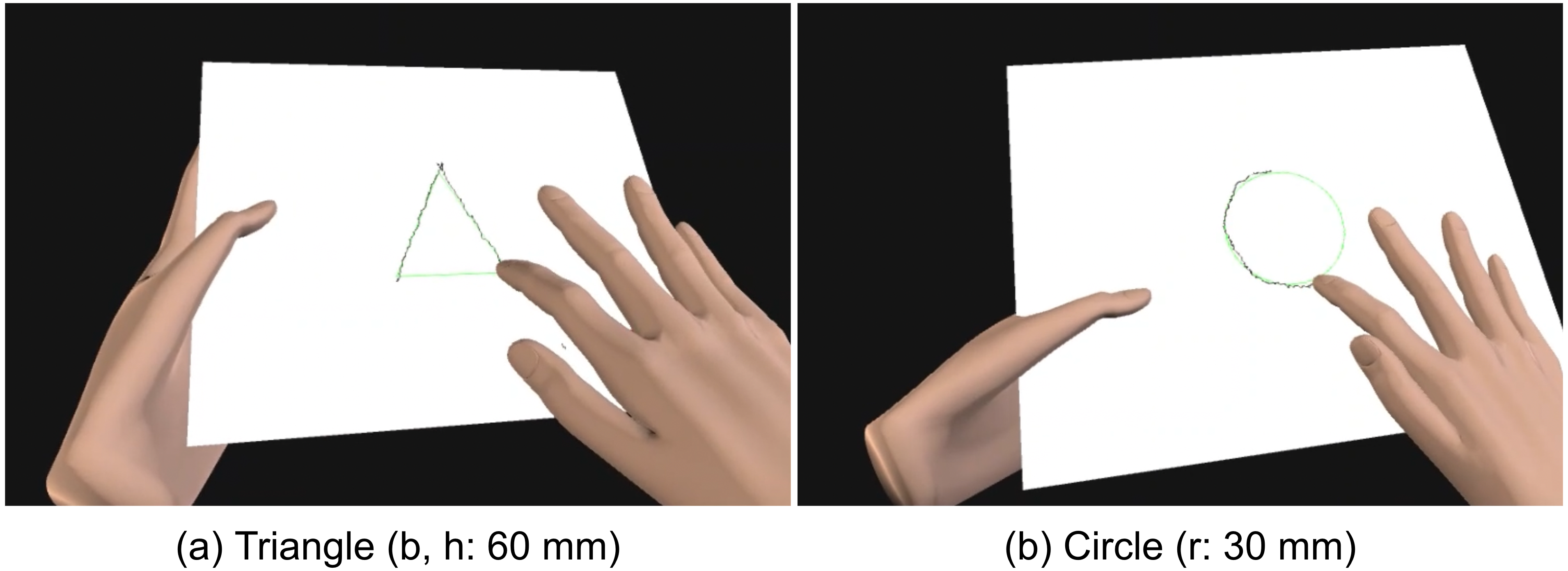}
\caption{Screenshots of the two tracing tasks performed in this study: (a) a triangle with a base length and height of 60 mm and (b) a circle with a radius of 30 mm.}
\vspace{-2pt}
\label{tracing_task}
\end{figure}

Besides discrete input, such as selection, on the virtual surface, we also examined continuous input interactions. Tracing has been widely studied in previous research on finger-based interaction for continuous input \cite{xiao18mrtouch, NormalTouch_and_TextureTouch}. We provided two target shapes for the tracing task: a triangle with a base length and height of 60 mm (Figure \ref{tracing_task}a) and a circle with a 30 mm radius (Figure \ref{tracing_task}b). Both target shapes were displayed at the center of the virtual surface. 

\subsubsection{Sketching}

\begin{figure}[t]
\centering
\includegraphics[width=\linewidth]{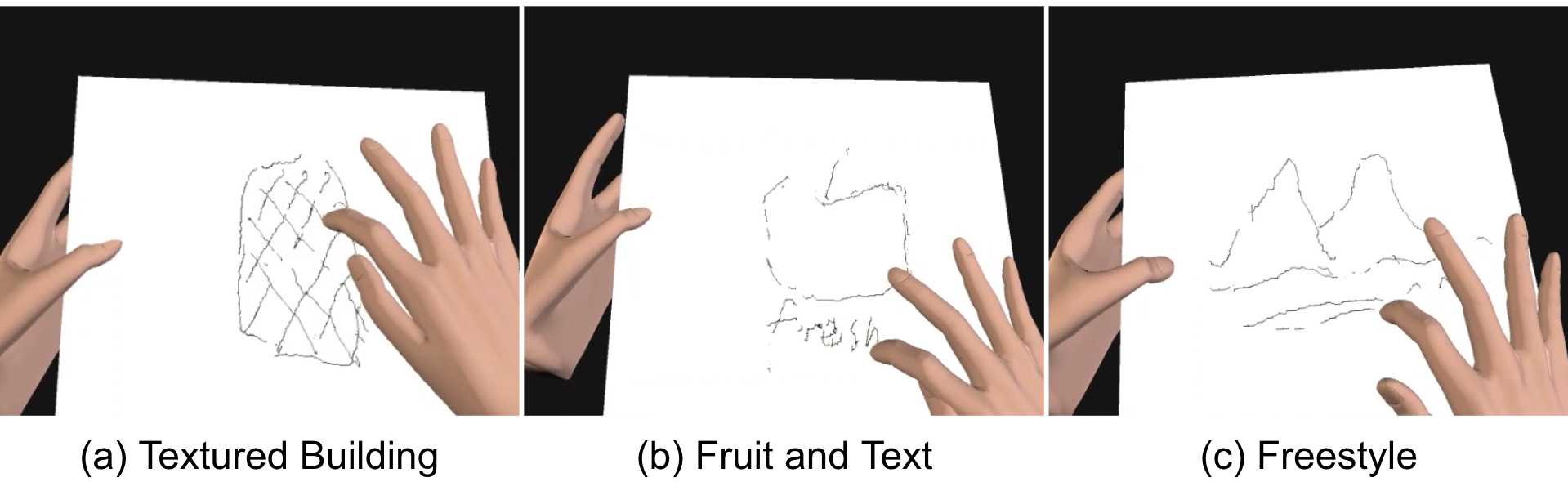}
\caption{Screenshots of the three sketching tasks performed in this study: (a) a textured building, (b) a fruit shape with text, and (c) a freestyle sketch.}
\vspace{-3pt}
\label{sketching_task}
\end{figure}

In the context of continuous input, sketching is a more advanced task than tracing, requiring more complex and intricate strokes to represent various drawing contents \cite{evaluation_sketching_Arora, investigating_sketching_Israel, finger_pen_stroke}. In our study, participants were asked to draw three sketches involving various shapes and textures: a textured building, a fruit shape with some text, and a freestyle sketch (Figure \ref{sketching_task}). For building- and fruit-sketching, two reference sketches \cite{SymbiosisSketch} were given. Participants were required to copy them freehand.

We instructed participants to complete the tracing and sketching tasks with as few smooth strokes as possible while sketching enough details.

\subsection{Participants}
Twelve individuals (six males and six females, 19-30 years old) participated in the study. All participants were right-handed and recruited from a university. Three participants were familiar with VR devices through VR games, and six participants had tried VR headsets before. All participants had non-expert sketching experience with paper and pen, but none had prior sketching experience in VR. The experiment lasted 35--50 minutes, and participants were paid USD 25. The study protocol was approved by the University of Virginia Institutional Review Board. 

\subsection{Study Design}
This study comprised three within-subject experiments, each focusing on a distinct interaction task and varying haptic feedback conditions. In the \textit{Selection} task, a $3\times2$ within-subject design was used with two independent variables (IVs): \textit{Haptic Condition} (\textit{Bare}, \textit{Tactile}, \textit{Physical}) and \textit{Target Size} (Small, Large). The \textit{Tracing} task also followed a $3\times2$ design with \textit{Haptic Condition} (\textit{Bare}, \textit{Tactile}, \textit{Physical}) and \textit{Target Shape} (Triangle, Circle) as IVs. Finally, the \textit{Sketching} task employed a $3\times3$ design with \textit{Haptic Condition} (\textit{Bare}, \textit{Tactile}, \textit{Physical}) and \textit{Sketch Type} (\textit{Building}, \textit{Fruit}, \textit{Freestyle}) as IVs.

The presentation of haptic feedback conditions was counterbalanced using a balanced Latin square design. The tasks were presented in a fixed sequence: selection, tracing, and sketching. Participants completed a $\sim3$ minute training block in each haptic condition for every task to familiarize themselves with the condition. In general, with each haptic feedback, participants performed 24 selection tasks for small targets, 24 selection tasks for large targets, five triangle-tracing tasks, five circle-tracing tasks, one building sketching, one fruit sketching, and one freestyle sketching task for a total of 183 trials (i.e., (48 (selection tasks) + 10 (tracing tasks) + 3 (sketching tasks)) x 3 (haptic feedback conditions) = 183 trials). 

\subsection{Evaluation Metrics}
\subsubsection{Data Collection}
During the experiment, a log file recorded the time and 3D positions of the hands, virtual surface, physical surface, and contact between the hand and surface at a rate of 60 Hz. The contact point was calculated by mapping the 3D position of the center of the right index fingertip onto the virtual surface. The mapping direction is the line of sight from the eye to the operating finger. Both tracing and sketching results were derived from the 2D texture of the virtual surface, as shown in Figure \ref{tracing_task} and Figure \ref{sketching_task}.

\subsubsection{Selection Task Performance}
Selection performance was evaluated by the Fitts' Law Throughput (TP) and the number of errors. A selection error occurred when the participant touched the surface, but the contact point was outside the target.

\subsubsection{Tracing Task Performance}
To evaluate the quality of tracings, we resampled each tracing trajectory to 100 equidistant points (px, py) on the 2D coordinate, with the center located at (0,0). We then calculated the tracing precision, stroke smoothness, stroke continuity, and task completion time.

\paragraph{Tracing precision (MPD)} This measure was calculated by Mean Projected Deviation (MPD) \cite{evaluation_sketching_Arora}, which represents the average distance between the points of the participant-drawn stroke and the target circle or triangle. A smaller offset distance indicates that the stroke the participant drew had higher precision and was more consistent with the target trajectory. 

To calculate the MPD for tracing a triangle, the tracing trajectory was first segmented into three straight lines corresponding to the three edges of the target triangle. Each line segment was then transformed to align with the x-axis, with its center positioned at (0,0). After this transformation, the MPD was determined as the average perpendicular deviation of the traced points from the target line (i.e., the x-axis). As shown in Formula \ref{MPD_triangle}, $MPD_{triangle}$ takes $n$ as the number of points on the stroke, $py_{i}$ represents the corresponding value of the $i^{th}$ point on the $y$ axis.

\begin{equation}
MPD_{triangle} = \frac{1}{n}\sum^{n}_{i=1}(py_{i})
\label{MPD_triangle}
\end{equation}

The circle MPD was calculated as shown in Formula \ref{MPD_circle}, where $n$ is the number of points on each trajectory, $px_{i}$ and $py_{i}$ are the values of the $i^{th}$ pixel on the $x$ and $y$ axes, and $r$ is the radius of the target. A smaller offset indicates that the stroke drawn by the participant had higher precision and was more consistent with the target trajectory.

\begin{equation}
MPD_{circle} = \frac{1}{n}\sum^{n}_{i=1}(\sqrt{(px_i)^2+(py_i)^2}-r)
\label{MPD_circle}
\end{equation}

\paragraph{Stroke smoothness (MFD)} We computed Mean Fairness Deviation (MFD) \cite{evaluation_sketching_Arora} to assess the average smoothness of strokes during one tracing trajectory. MFD was defined as the mean curvature difference between consecutive points on a stroke, with smaller MFD values indicating smoother strokes. As shown in Equation \ref{MFD}, $n$ represents the number of points on a stroke, and $C_{i}$ is the curvature at the $i^{th}$ point.

\begin{equation}
MFD = \frac{1}{n-1}\sum^{n-1}_{i=1}\frac{|C_{i}-C_{i+1}|}{avg\left(C_{i},C_{i+1}\right)}
\label{MFD}
\end{equation}

\paragraph{Stroke Continuity (Breakpoints)} Since participants could complete a single tracing trial using multiple strokes, continuity was evaluated based on the number of breakpoints within a trial. A breakpoint was defined as a point where two strokes, initiated by separate touches, failed to meet at their endpoints. Such discontinuities can negatively impact the perceived quality of a stroke intended to be continuous \cite{wiese2010invest_free_hand_sketching}. Therefore, a higher number of breakpoints indicates a deficiency in stroke continuity.

\paragraph{Task Completion Time (CT)} CT was calculated from the time participants started drawing the first stroke to the time they finished the last stroke.

\subsubsection{Sketching Task Performance}
To subjectively evaluate the quality of sketches, we recruited 30 reviewers from Amazon Mechanical Turk (MTurk) to rate three key attributes: stroke smoothness, stroke continuity, and sketch clarity, using a 10-point rating scale (1 being worst and 10 being best). Stroke smoothness refers to the consistency of stroke curvature \cite{evaluation_sketching_Arora}; stroke continuity refers to the alignment of two strokes when they are intended to connect \cite{wiese2010invest_free_hand_sketching}; and sketch clarity refers to the overall recognizability of the sketch components (e.g., shapes, textures). Examples of smoothness, continuity, and clarity ratings are shown in \ref{appendix:sketching_rating_criteria}. In addition, we provided a collection of sketches from the three haptic conditions and asked reviewers to rank them based on overall sketch quality. Workers who participated in the study received USD 6 as compensation for their efforts.

\subsubsection{Learning Effects}
While the experiment used a counterbalanced presentation of interaction conditions, participants may still exhibit learning over time. To assess such effects, we analyzed performance trends across haptic condition orders to determine whether practice influenced selection and tracing performance.

\subsubsection{Bimanual Behavior}
The dominant hand and non-dominant hands had different roles in this study, with the non-dominant hand holding and stabilizing the surface while the dominant hand interacted with it. To investigate this bimanual behavior, we recorded and analyzed the involvement and coordination of both hands during each selection, tracing, and sketching trial. The \textit{Dominant Hand Trajectory} along the surface's normal direction was visualized to understand its behavior at the start of touch (i.e., before contact). Specifically, the distance between the dominant finger and the surface was measured within 1 second before contact. The time frame was chosen based on the typical screen touch time ($\sim$ 0.5 s \cite{Zero_Latency_Tapping}) and the average contact time in our experiment ($\sim$ 0.7 s). Additionally, we calculated the \textit{Total Dominant Hand Movement} and \textit{Total Non-Dominant Hand Movement} (i.e., surface movement) to assess overall hand movement and hand coordination during tasks. 

\subsubsection{Questionnaires}
One standardized and one customized questionnaire were used to investigate the qualitative effects of haptic feedback on task performance. The first questionnaire adopted the NASA Task Load Index (NASA-TLX \cite{Colligan2015CognitiveWC}), which provided a standardized measure of workload when performing tasks with different haptic feedback. Participants completed the NASA-TLX survey after each condition. The second questionnaire was customized to collect participants' feedback on interaction confidence, fatigue, and the perceived quality of different haptic feedback methods. 

\subsection{Data Analysis}
For parametric measurements, all data were tested for normality. Repeated-measure ANOVA tests were used to analyze the normally distributed data with $p = 0.05$ as the significance level. If further post-hoc testing was necessary, pairwise t-tests with Bonferroni correction were conducted. For the non-normally distributed data, Applied Aligned Rank Transform (ART) \cite{wobbrock2011ART} tests were adopted, and the ART-C \cite{elkin_ARTC} procedure was used for pairwise post-hoc comparisons when necessary. Regarding the non-parametric measures, a Friedman's ANOVA test was performed on the Likert results. Mann-Whitney post-hoc tests with a Bonferroni correction were conducted if needed.

%% file: Results.tex
\section{RESULTS}
This study evaluated participants' performance of selecting, tracing, and sketching tasks while interacting with the three different haptic conditions: \textit{no haptic feedback (Bare)}, \textit{tactile feedback (Tactile)}, and \textit{physical surface (Physical)}. The outcomes of each task are shown and explained in turn. 

\subsection{Selection Performance}

\subsubsection{Selection Errors}

\begin{figure}[t]
\centering
\includegraphics[width=\linewidth]{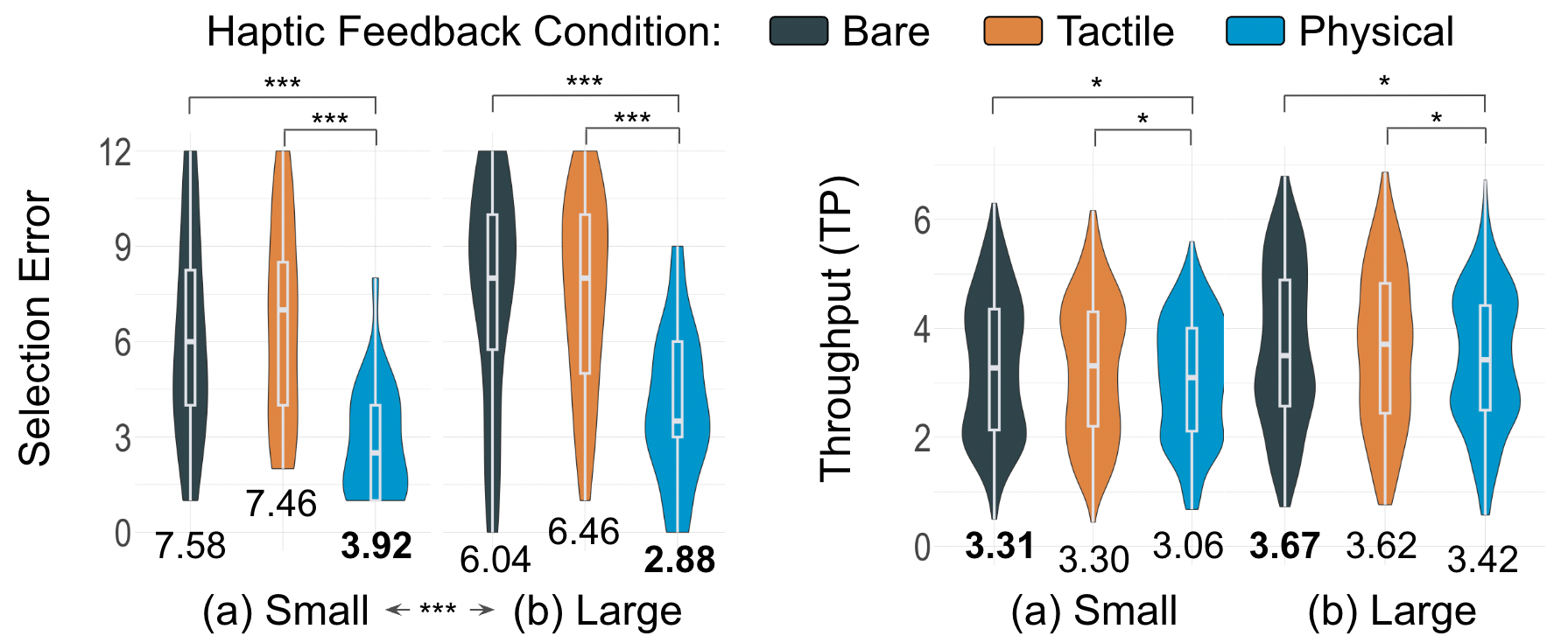}
\caption{Number of errors and throughput under the three haptic conditions. Mean values were shown at the bottom of each condition.}
\vspace{-3pt}
\label{select_error_TP}
\end{figure}

Both the haptic condition ($F_{(2,22)}=40.1,\ p<0.001$) and the target size ($F_{(1,11)}=27.0,\ p<0.001$) showed a significant effect on selection errors. As shown in Figure \ref{select_error_TP}, participants made the fewest errors under the \textit{Physical} condition, with a mean error of $3.92\ (SD = 2.32)$ for small targets and $2.88\ (SD = 1.80)$ for large targets. This result was followed by the \textit{Bare} condition, with a mean error of $7.58\ (SD = 3.50)$ for small targets and $6.04\ (SD = 3.26)$ for large targets, and the \textit{Tactile}, with a mean error of $7.46\ (SD = 3.04)$ for small targets and $6.46\ (SD = 3.15)$ for large targets. Overall, having a physical surface significantly reduced selection errors by half compared to the \textit{Bare} ($p<0.001$) and \textit{Tactile} ($p<0.001$) conditions.

\subsubsection{Throughput (TP)}
However, \textit{Physical} did not show an advantage on TP ($F_{(2,22)}=27.0,\ p<0.001$) as shown in Figure \ref{select_error_TP}. \textit{Bare} resulted in the highest TP for both small ($M=3.31,\ SD=1.30$) and large targets ($M=3.67,\ SD=1.47$), followed by \textit{Tactile}, with a mean TP of $3.30\ (SD = 1.30)$ for small targets and $3.62\ (SD = 1.45)$ for large targets, and \textit{Physical}, with a mean TP of $3.06\ (SD = 1.11)$ for small targets and $3.42\ (SD = 1.26)$ for large targets. Significant differences were found across all haptic conditions ($p<0.05$) except between \textit{Bare} and \textit{Tactile} conditions ($p>0.05$). 

\subsubsection{Fitts’ Law MT/ID relationship}

\begin{figure}[t]
\centering
\includegraphics[width=\linewidth]{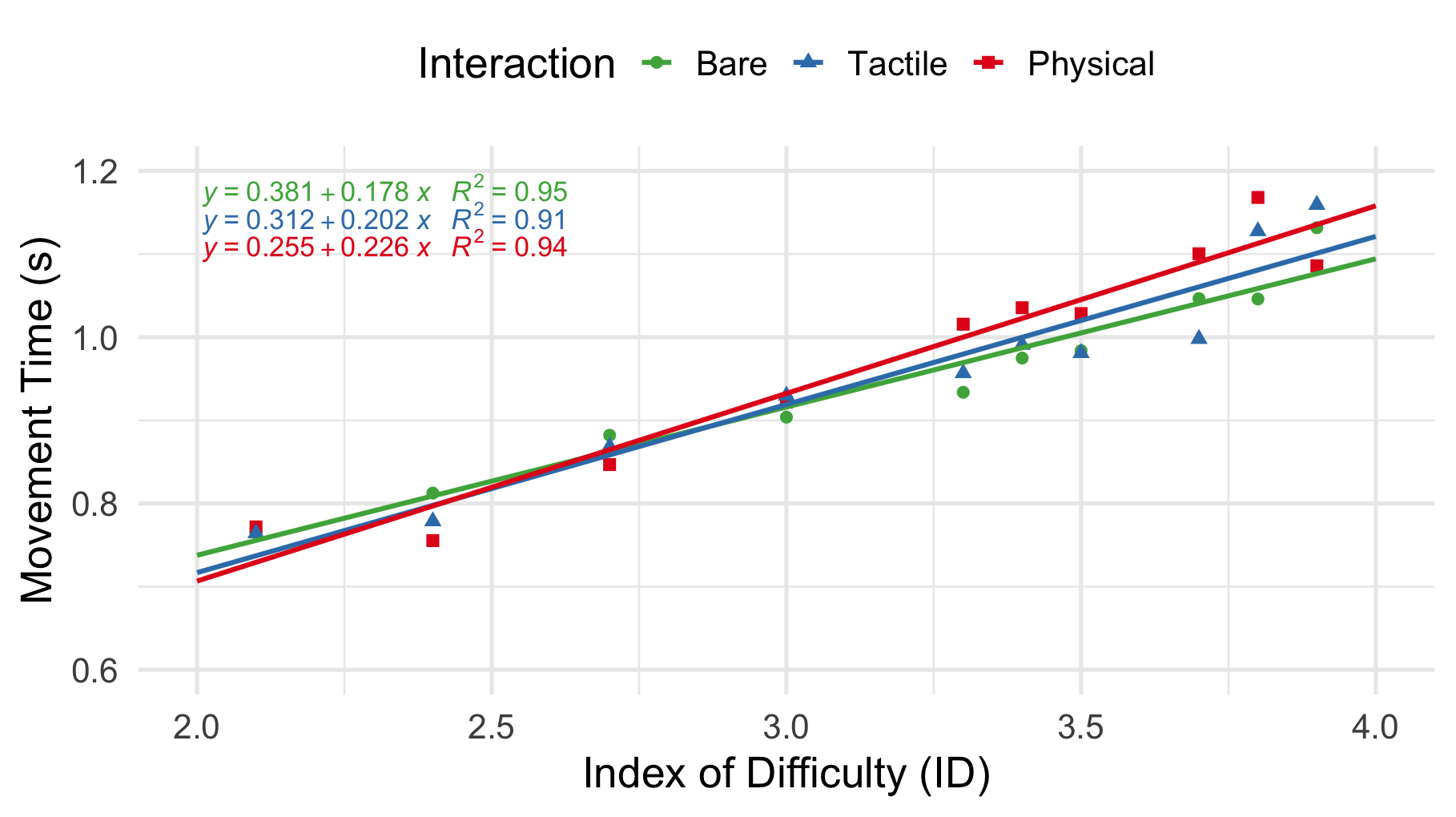}
\caption{The relationship between Fitts' Law Movement Time (MT) and Index of Difficulty (ID) during the selection task for the three haptic feedback conditions.}
\label{fittslawMT_ID}
\end{figure}

A linear regression of MT and ID \cite{bi2013ffitts} was performed to examine the goodness of fit. For each ID, the mean MT under different haptic feedback conditions was calculated. According to the $R^2$ value (Figure \ref{fittslawMT_ID}), the prediction of selection performance under all conditions was acceptable. \textit{Bare} generated the smallest slope, followed by \textit{Tactile} and \textit{Physical}, indicating lower sensitivity of MT to task difficulty.

\subsubsection{Learning Effects}

A two-way repeated-measures ANOVA was conducted to examine learning effects across haptic condition orders. For both small and large targets, the number of errors was significantly affected only by haptic condition (small: $F_{(2,16)}=17.28,\ p<0.001$; large: $F_{(2,16)}=10.10,\ p<0.001$), but not by order (small: $F_{(2,16)}=17.28,\ p<0.001$; large: $F_{(2,16)}=1.18,\ p>0.05$).

\begin{figure}[t]
\centering
\includegraphics[width=\linewidth]{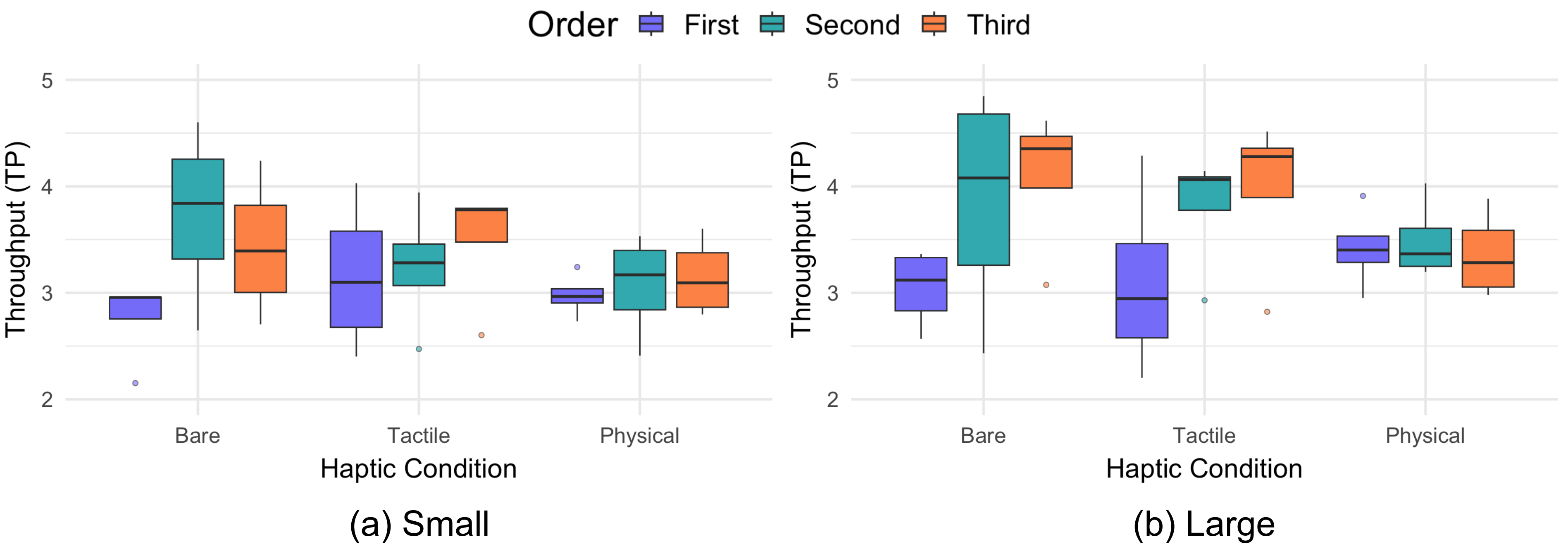}
\caption{Learning effects on TP across interaction conditions during selection tasks.}
\vspace{2pt}
\label{select_TP_learning}
\end{figure}

Regarding TP, for both small-target and large-target selection as shown in Figure \ref{select_TP_learning}, order had a significant effect on TP (small: $F_{(2,16)}=4.37,\ p<0.05$; large: $F_{(2,16)}=4.00,\ p<0.05$). Post-hoc comparisons found that TP was significantly lower when the haptic condition was performed first ($p<0.05$).

These results indicate that while selection errors did not vary systematically across blocks, TP changed with the order of haptic conditions, consistent with motor familiarization, as greater practice tended to increase TP.

\subsection{Bimanual Behavior in the Selection Task}

\subsubsection{Dominant Hand Trajectory}

\begin{figure}[t]
\centering
\includegraphics[width=\linewidth]{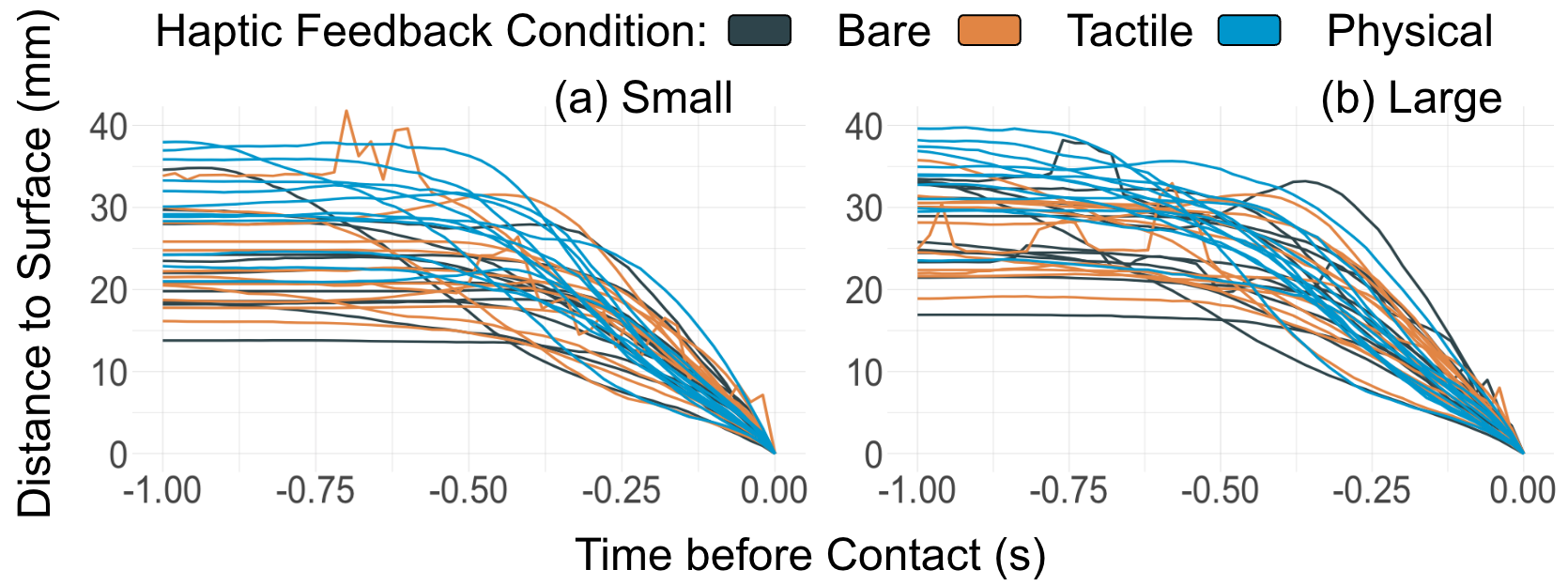}
\caption{ The dominant hand trajectory along the surface’s normal direction within one second before contact. The selection task used two target sizes: (a) small and (b) large.}
\label{select_finger_surface_dis}
\end{figure}

The dominant hand trajectory along the surface's normal direction (i.e., finger-surface distance) before contact is shown in Figure \ref{select_finger_surface_dis}. Both the haptic condition ($F_{(2,22)}=11.9,\ p<0.001$) and the target size ($F_{(1,11)}=6.6,\ p<0.05$) significantly influenced this trajectory. The most pronounced difference in finger-surface distance was observed at $0.75$ s before contact during both target selections. The \textit{Physical} condition resulted in a significantly larger finger-surface distance compared to the \textit{Tactile} and \textit{Bare} conditions ($p<0.001$). Under the \textit{Physical} condition, participants held their dominant finger $29.9\text{ mm}$ from the surface for small targets and $32.4\text{ mm}$ for large targets. In comparison, the \textit{Tactile} condition averaged $23.0\text{ mm}$ and $26.4\text{ mm}$, while the \textit{Bare} condition averaged $22.2\text{ mm}$ and $26.8\text{ mm}$ for small and large targets, respectively. This result may indicate that under the \textit{Physical} condition, most participants initiated touch by holding their dominant finger farther away from the surface and then approaching it at a higher speed. 

According to Figure \ref{select_finger_surface_dis}, the hand trajectory under the \textit{Tactile} and \textit{Bare} conditions exhibited more ``bumps'', suggesting that the dominant finger made more small up-and-down movements when selecting on the surface. These movements could have been caused by unintended hand shaking in the air or deliberate hand adjustments.

\subsubsection{Total Dominant Hand Movement}

\begin{figure}[t]
\centering
\includegraphics[width=\linewidth]{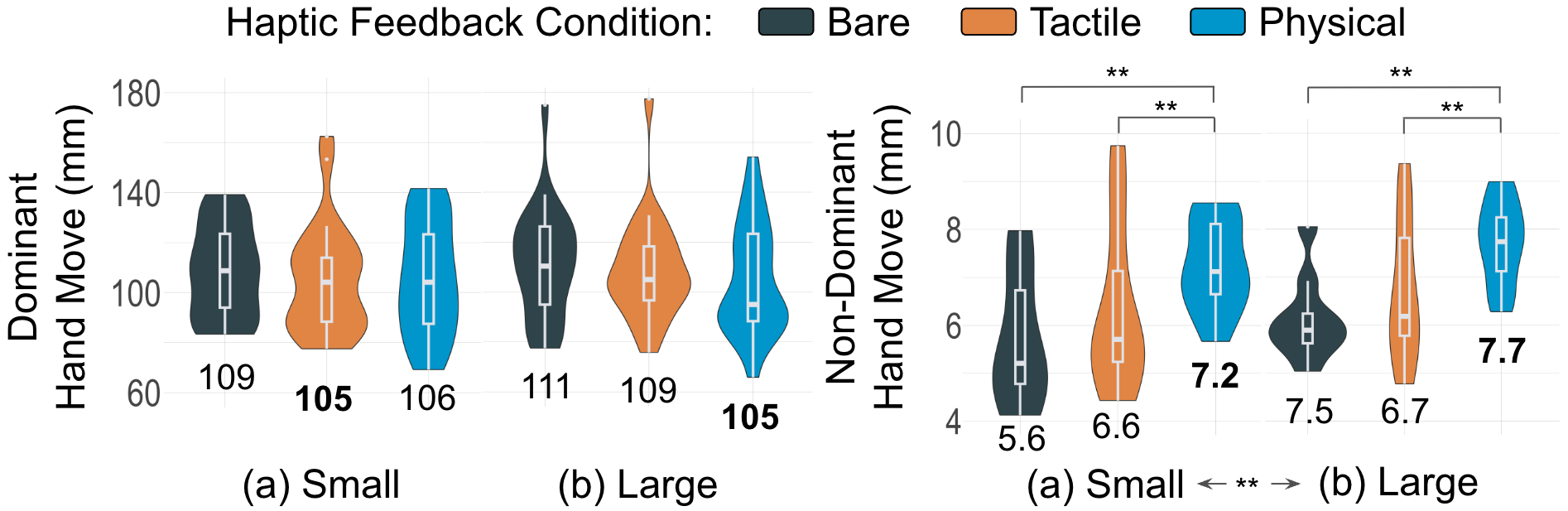}
\caption{ The total dominant hand movement and non-dominant hand movement when selecting one (a) small target and (b) large target under different conditions. Mean values were shown at the bottom of each haptic condition. }
\label{select_hand_surface_move}
\end{figure}

No significant differences were observed in total dominant hand movement among the three haptic conditions $(F_{(2,22)}=0.79,\ p>0.05)$ or between the two target sizes $(F_{(1,11)}=0.29,\ p>0.05)$ during the selection task. As shown in Figure \ref{select_hand_surface_move}, the dominant hand movement across all conditions was approximately $110\ mm$, including both movements along the surface normal and travel between consecutive targets.

\subsubsection{Total Non-Dominant Hand Movement}
Haptic condition significantly influenced total non-dominant hand movement (i.e., surface movement) during selection $(F_{(2,22)}=7.89,\ p<0.001)$, and target size also showed a significant effect $(F_{(1,11)}=8.41,\ p<0.01)$.

Figure \ref{select_hand_surface_move} shows that selecting on the physical surface resulted in the greatest non-dominant hand movement, averaging $7.24\ (SD=0.94)\ mm$ for small-target selection and $7.72\ (SD=0.88)\ mm$ for large-target selection. In contrast, under the \textit{Tactile} condition, the mean non-dominant hand movement was $6.64\ (SD=2.63)\ mm$ for small targets and $6.71\ (SD=1.49)\ mm$ for large targets. The \textit{Bare} condition resulted in the least non-dominant hand movement, with participants moving the surface $5.57\ (SD=1.42)\ mm$ for small-target selection and $7.47\ (SD=2.56)\ mm$ for large-target selection.

This finding may suggest that the physical surface increased the involvement and coordination of the non-dominant hand, as the surface was actively held and manipulated in the air during the task.

\subsection{Tracing Performance}

\begin{figure}[t]
\centering
\includegraphics[width=\linewidth]{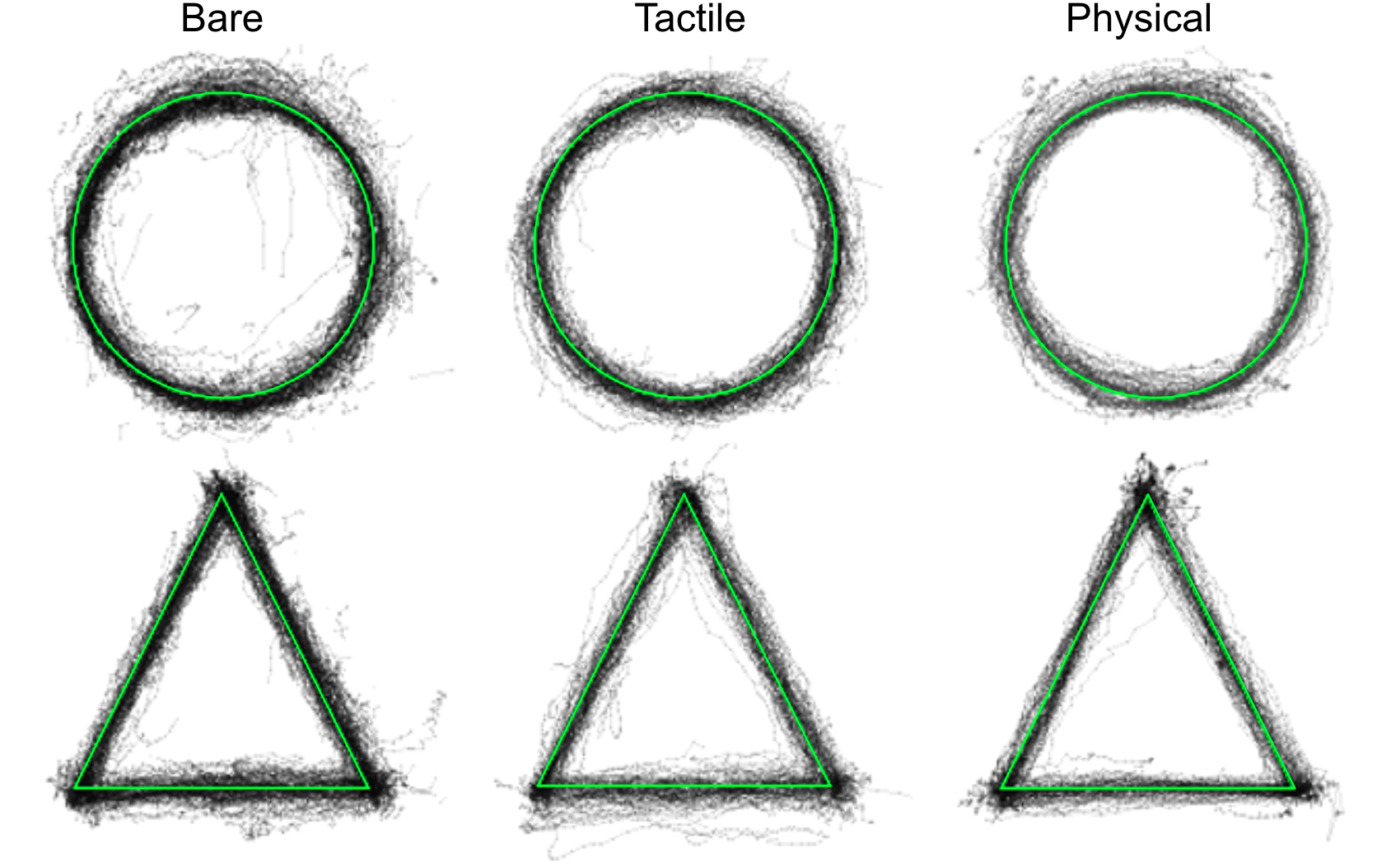}
\caption{The tracing strokes drawn by all participants are depicted in black. The target trajectories are highlighted in green. }
\label{tracking_trends}
\end{figure}

Figure \ref{tracking_trends} illustrates the collection of raw strokes drawn by all participants. The impact of the haptic conditions and tracing shapes on various performance metrics is subsequently analyzed in detail.

\subsubsection{Tracing Precision (MPD)}

\begin{figure}[t]
\centering
\includegraphics[width=\linewidth]{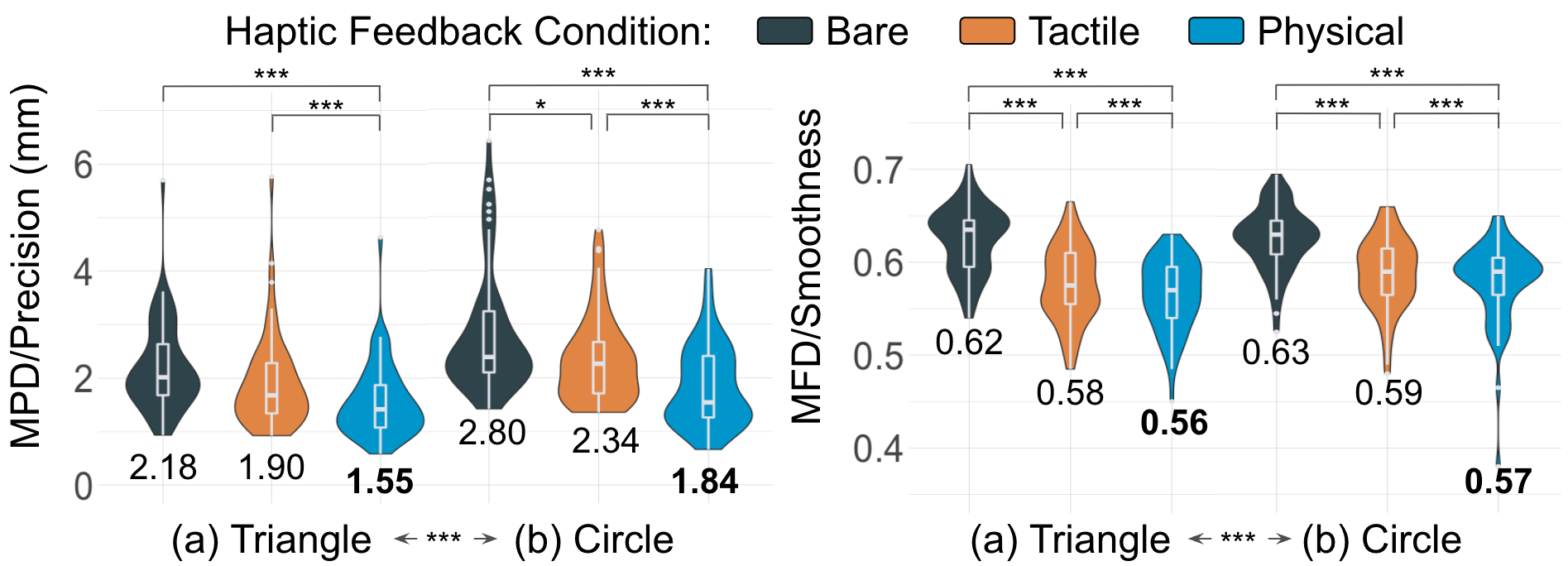}
\caption{ Mean Projected Deviation (MPD: precision) and Mean Fairness Deviation (MFD: smoothness) of the tracing strokes. The tracing task used two target shapes: (a) triangle and (b) circle. Mean values were shown at the bottom of each haptic condition.}
\label{tracing_MPD_MFD}
\end{figure}

As shown in Figure \ref{tracing_MPD_MFD}, both haptic conditions ($F_{(2,22)}=46.6,\ p<0.001$) and target shape ($F_{(1,11)}=31.5,\ p<0.001$) had significant effects on tracing precision, with lower MPD values indicating higher precision. The physical surface significantly improved triangle-tracing precision ($p<0.001$), while no significant difference was found between the \textit{Tactile} and \textit{Bare} conditions ($p>0.05$). Tracing a triangle with a physical surface ($M=1.55,\ SD=0.68$ mm) reduced MPD by 18.4\% compared to \textit{Tactile} ($M=1.90,\ SD=0.85$ mm) and by 28.9\% compared to \textit{Bare} ($M=2.18,\ SD=0.81$ mm). For circle tracing, using a physical surface ($M=1.84,\ SD=0.75$ mm) reduced MPD by 21.4\% compared to \textit{Tactile} ($M=2.34,\ SD=0.81$ mm) and by 16.4\% compared to \textit{Bare} ($M=2.80,\ SD=1.14$ mm), with all differences across haptic conditions being statistically significant ($p<0.05$). 

An interaction was found between target shape and haptic condition ($F_{(2,22)}=11.1,\ p<0.001$). Tracing a circle ($M=4.65,\ SD=1.97$ mm) produced a larger MPD than tracing a triangle ($M=1.88,\ SD=0.82$ mm), particularly in the \textit{Bare} and \textit{Tactile} conditions ($p<0.01$). While using physical surfaces improved drawing precision, the benefits appeared greater for tasks involving curved lines, such as circle tracing.

\subsubsection{Stroke Smoothness (MFD)}
Statistical analysis revealed that haptic condition had a significant influence on the smoothness of tracing strokes ($F_{(2,22)}=63.6,\ p<0.001$; Figure \ref{tracing_MPD_MFD}), with lower MFD values indicating greater smoothness. Strokes drawn under the \textit{Physical} condition were the smoothest (Figure \ref{tracing_MPD_MFD}), with an average MFD of $0.56\ (SD=0.03)$ for triangles and $0.57\ (SD=0.04)$ for circles. The \textit{Tactile} condition followed, showing an average MFD of $0.58\ (SD=0.04)$ for triangles and $0.59\ (SD=0.04)$ for circles, while the \textit{Bare} condition showed the least smooth strokes with $0.62\ (SD=0.04)$ for triangles and $0.63\ (SD=0.03)$ for circles. Post-hoc analyses revealed that the differences across the three haptic conditions were all significant ($p<0.001$). 

Additionally, target shape had a significant effect on stroke smoothness ($F_{(1,11)}=8.32,\ p<0.01$), with an interaction found between haptic condition and target shape ($F_{(2,22)}=3.28,\ p<0.05$). Overall, tracing a triangle ($M=0.59,\ SD=0.05$) produced smoother strokes compared to tracing a circle ($M=0.60,\ SD=0.04$).

\subsubsection{Stroke Continuity (Breakpoints)}

\begin{figure}[t]
\centering
\includegraphics[width=\linewidth]{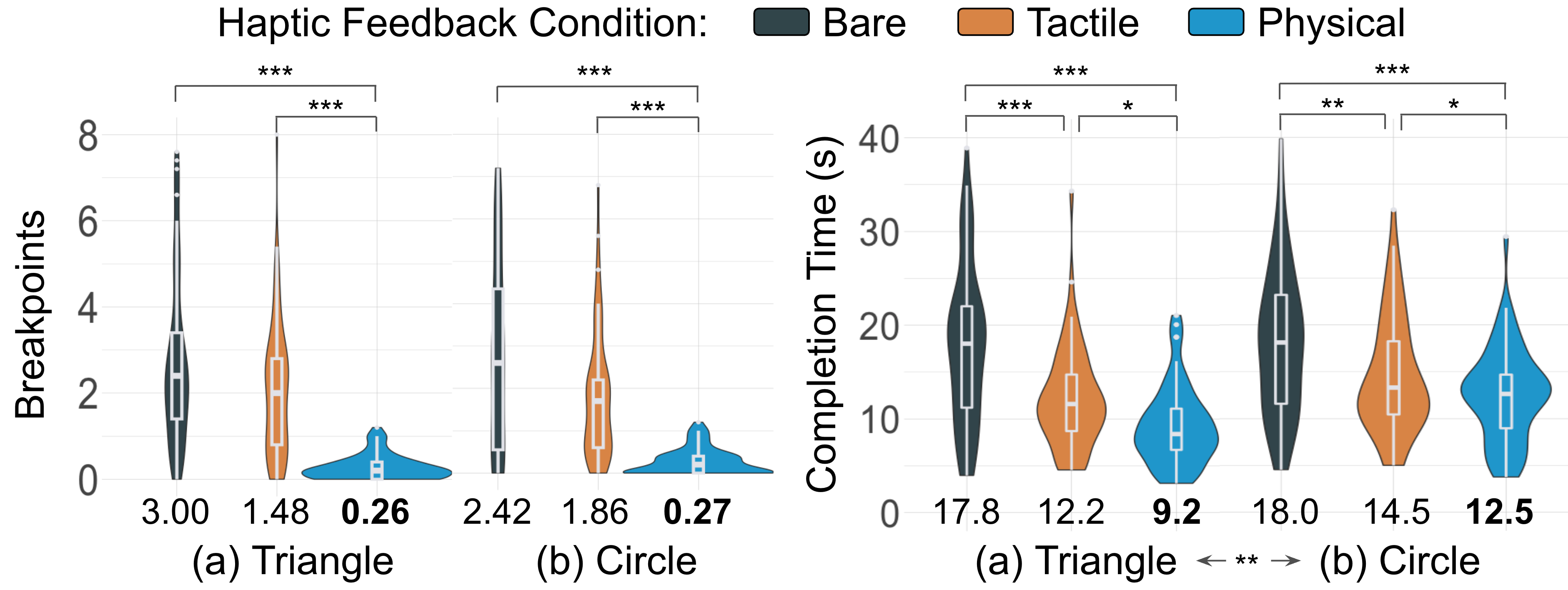}
\caption{Number of breakpoints (continuity) and task completion time when tracing a (a) triangle and (b) circle. Mean values were shown at the bottom of each haptic condition.}
\label{tracing_breakpoint_CT}
\end{figure}

Stroke continuity was significantly affected by haptic condition ($F_{(2,22)}=40.07,\ p<0.001$). As shown in Figure \ref{tracing_breakpoint_CT}, participants produced significantly more continuous strokes when tracing on the physical surface than in the \textit{Bare} and \textit{Tactile} conditions ($p<0.001$). A triangle trajectory on the physical surface had an average of $0.26\ (SD=0.31)$ breakpoints, while a circle trajectory had an average of $0.27\ (SD=0.29)$ breakpoints. The \textit{Tactile} condition showed more breakpoints, averaging $1.83\ (SD=1.48)$ for triangles and $2.21\ (SD=1.86)$ for circles. The \textit{Bare} condition had the worst stroke continuity, where participants made $2.96\ (SD=2.34)$ breakpoints when tracing a triangle and $3.06\ (SD=2.42)$ breakpoints when tracing a circle. 

\subsubsection{Completion Time (CT)}
CT was significantly affected by haptic condition ($F_{(2,22)}=12.1,\ p<0.001$) and target shape ($F_{(1,11)}=10.9,\ p<0.01$) as shown in Figure \ref{tracing_breakpoint_CT}. Tracing a triangle on the physical surface ($M=9.20,\ SD=4.18$ s) was 24.3\% faster than with \textit{Tactile} feedback ($M=12.16,\ SD=5.37$ s) and 36.4\% faster than with \textit{Bare} feedback ($M=17.80,\ SD=8.37$ s). For circle tracing, the \textit{Physical} condition ($M=12.53,\ SD=4.98$ s) was 13.4\% faster than \textit{Tactile} ($M=14.47,\ SD=6.19$ s) and 30.4\% faster than \textit{Bare} ($M=18.00,\ SD=7.76$ s). Post-hoc tests confirmed that all differences in CT across haptic conditions were statistically significant ($p < 0.05$).  

Tracing triangles ($M=13.1,\ SD=7.1$ s) was faster than tracing circles ($M=15.0,\ SD=7.1$ s). A significant interaction between haptic condition and target shape ($F_{(2,22)}=3.54,\ p<0.05$) further suggests that triangle tracing benefited more from having a physical surface ($p<0.001$).

\subsubsection{Learning Effects}

For both triangle and circle tracing, MPD was significantly affected by haptic condition (triangle: $F_{(2,16)}=7.53,\ p<0.01$; circle: $F_{(2,16)}=8.65,\ p<0.01$), but not by order (triangle: $F_{(2,16)}=2.08,\ p>0.05$; circle: $F_{(2,16)}=0.28,\ p>0.05$). Similarly, the analysis showed a significant effect of haptic condition on CT (triangle: $F_{(2,16)}=10.64,\ p<0.001$; circle: $F_{(2,16)}=6.28,\ p<0.01$), but no effect of order (triangle: $F_{(2,16)}=0.78,\ p>0.05$; circle: $F_{(2,16)}=0.15,\ p>0.05$). These results indicate that differences in tracing precision and speed are not driven by order, with no evidence of learning effects.

\subsection{Bimanual Behavior in the Tracing Task}

\subsubsection{Dominant Hand Trajectory}

\begin{figure}[t]
\centering
\includegraphics[width=\linewidth]{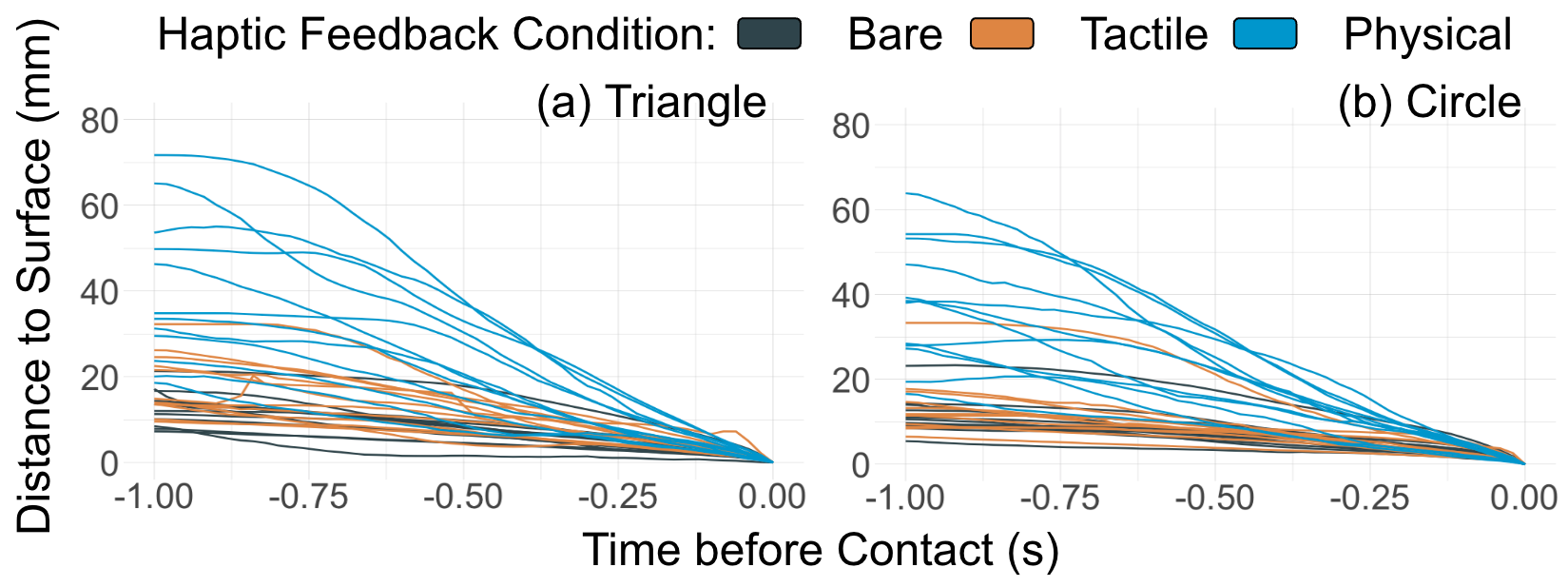}
\caption{ The dominant hand trajectory along the surface’s normal direction (i.e.,
finger-surface distance) within one second before contact. The tracing task used two target shapes: (a) triangle and (b) circle.}
\label{trace_finger_surface_dis}
\end{figure}

Figure \ref{trace_finger_surface_dis} illustrates the dominant hand trajectory along the surface's normal direction (i.e., finger-surface distance) before contact. The haptic condition significantly affected this distance $(F_{(2,22)}=28.2.5,\ p<0.001)$. Under the \textit{Physical} condition, most participants initiated touch by holding their dominant finger significantly farther away from the surface, then approaching the surface with a higher speed compared to \textit{Tactile} and \textit{Bare} conditions ($p<0.001$). The most significant difference in finger-surface distance was observed at $0.75$ s before contact during triangle and circle tracing. 

Under the \textit{Physical} condition, participants held their dominant finger $32.0\ (SD=18.3)$ mm from the surface during triangle tracing and $30.6\ (SD=14.2)$ mm during circle tracing. This distance was 150\% significantly greater than in the \textit{Tactile} and \textit{Bare} conditions ($p<0.05$). The \textit{Tactile} condition showed mean finger-surface distances of $15.8\ (SD=6.83)$ mm for triangles and $12.1\ (SD=6.81)$ mm for circles, while the \textit{Bare} condition showed $13.0\ (SD=7.60)$ mm and $11.7\ (SD=7.09)$ mm, respectively.

\subsubsection{Total Dominant Hand Movement}

The haptic condition had a significant impact on the total dominant hand movement for tracing $(F_{(2,22)}=14.1,\ p<0.001)$. The ideal movement distance should approximate the circumference of the target circle ($\sim188.8\ mm$) or the perimeter of the target triangle ($\sim207.8\ mm$). Compared to these values, the \textit{physical} condition potentially supported the most efficient dominant hand movement with the closest travel distances as observed in Figure \ref{trace_hand_surface_move}. With a physical surface, participants moved the dominant finger $261.1\ (SD=40.1)\ mm$ when tracing a triangle and $280.4\ (SD=50.9)\ mm$ when tracing a circle. The \textit{Tactile} condition showed greater movement, averaging $290.1\ (SD=61.7)\ mm$ for triangles and $303.2\ (SD=79.3)\ mm$ for circles. The \textit{Bare} condition resulted in the largest movements, with mean distances of $380.2\ (SD=97.6)\ mm$ for triangles and $393.2\ (SD=98.4)\ mm$ for circles. Significant differences were found in all haptic conditions through post-hoc tests ($p<0.001$). 

In addition, the target shape significantly affected dominant hand movement during tracing $(F_{(1,11)}=4.28,\ p<0.05)$. Tracing a triangle resulted in less movement, with a mean distance of $310.6\ (SD=86.7)\ mm$ while tracing a circle had a mean value of $325.0\ (SD=92.0)\ mm$. This difference likely reflects the inherent difference in the ideal trajectory length of the two shapes (i.e., their circumferences).

\subsubsection{Total Non-Dominant Hand Movement}

\begin{figure}[t]
\centering
\includegraphics[width=\linewidth]{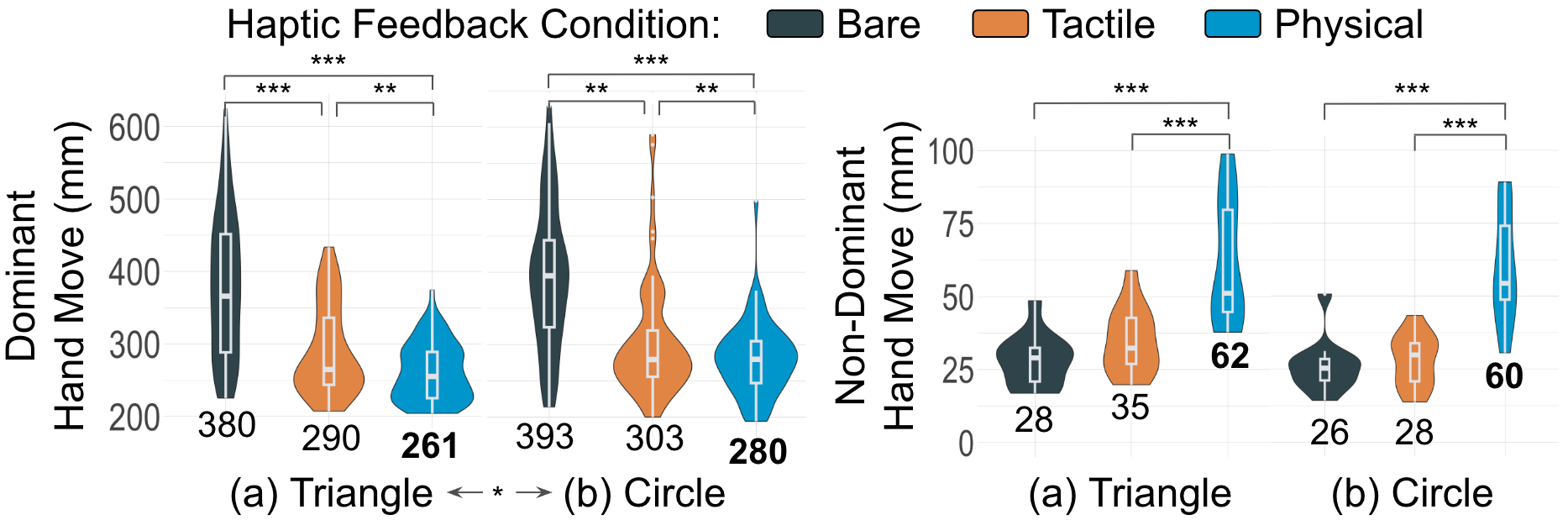}
\caption{ The total dominant hand movement and non-dominant hand movement when tracing one (a) triangle and (b) circle under different conditions. Mean values were shown at the bottom of each haptic condition.}
\label{trace_hand_surface_move}
\end{figure}

The haptic condition significantly influenced total non-dominant hand movement (i.e., surface movement) during tracing $(F_{(2,22)}=45.0,\ p<0.001)$. Tracing on the physical surface resulted in the greatest non-dominant hand movement, as shown in Figure \ref{trace_hand_surface_move}. Under the \textit{Physical} condition, the average total non-dominant hand movement distance was $61.9\ (SD=21.5)\ mm$ for triangle tracing and $60.1\ (SD=19.5)\ mm$ for circle tracing. In contrast, under the \textit{Tactile} condition, the distance was $34.8\ (SD=11.5)\ mm$ for triangle tracing and $27.9\ (SD=8.99)\ mm$ for circle tracing. The least non-dominant hand movement was observed under the \textit{Bare} condition, with participants moving the non-dominant hand $28.2\ (SD=9.08)\ mm$ when tracing a triangle and $26.0\ (SD=9.23)\ mm$ when tracing a circle. Overall, the \textit{Physical} condition resulted in twice the non-dominant hand movement compared to the \textit{Tactile} and \textit{Bare} conditions, with the difference being statistically significant $(p<0.001)$.

\begin{figure*}[t]
\centering
\includegraphics[width=0.9\textwidth, height=0.34\textheight]{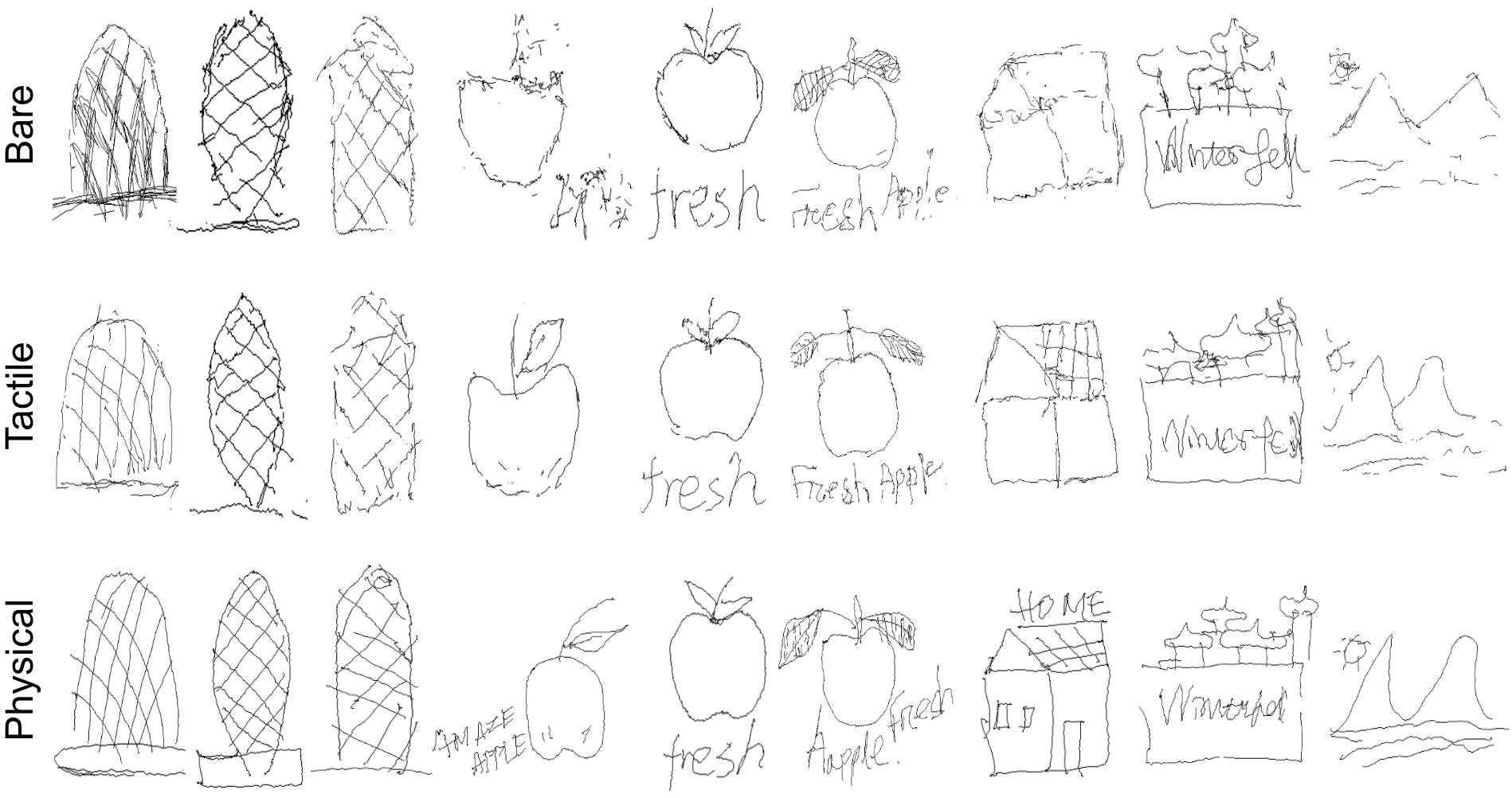}
\caption{The collection of sketches from three participants under the three haptic conditions.}
\vspace{-5pt}
\label{sketching_collection}
\end{figure*}

\subsection{Sketching Performance} 
Figure \ref{sketching_collection} presents example sketches from three participants, showcasing buildings, fruits, and freestyle content. We presented a collection of sketches from 12 participants to 30 reviewers for evaluation, and the sketches from each participant were randomized across the three conditions. 90\% of participants ranked the \textit{Physical} condition as producing the best overall sketch quality, 76.7\% of reviewers rated the \textit{Tactile} condition as providing medium-quality sketches, and 73.3\% of reviewers identified the \textit{Bare} condition as resulting in the lowest sketch quality. Detailed ratings for stroke smoothness, stroke continuity, and overall sketch clarity under different haptic conditions are presented next. To better illustrate the three criteria used to evaluate the sketches, we present examples from three participants in the (1) textured-building (Figure \ref{building_rating_exp}), (2) fruit-shape (Figure \ref{fruit_rating_exp}), and (3) freestyle (Figure \ref{freestyle_rating_exp}) sketching tasks, along with external reviewers’ ratings of stroke smoothness, stroke continuity, and overall clarity.

\subsubsection{Stroke Smoothness}

\begin{figure}[t]
\centering
\includegraphics[width=\linewidth]{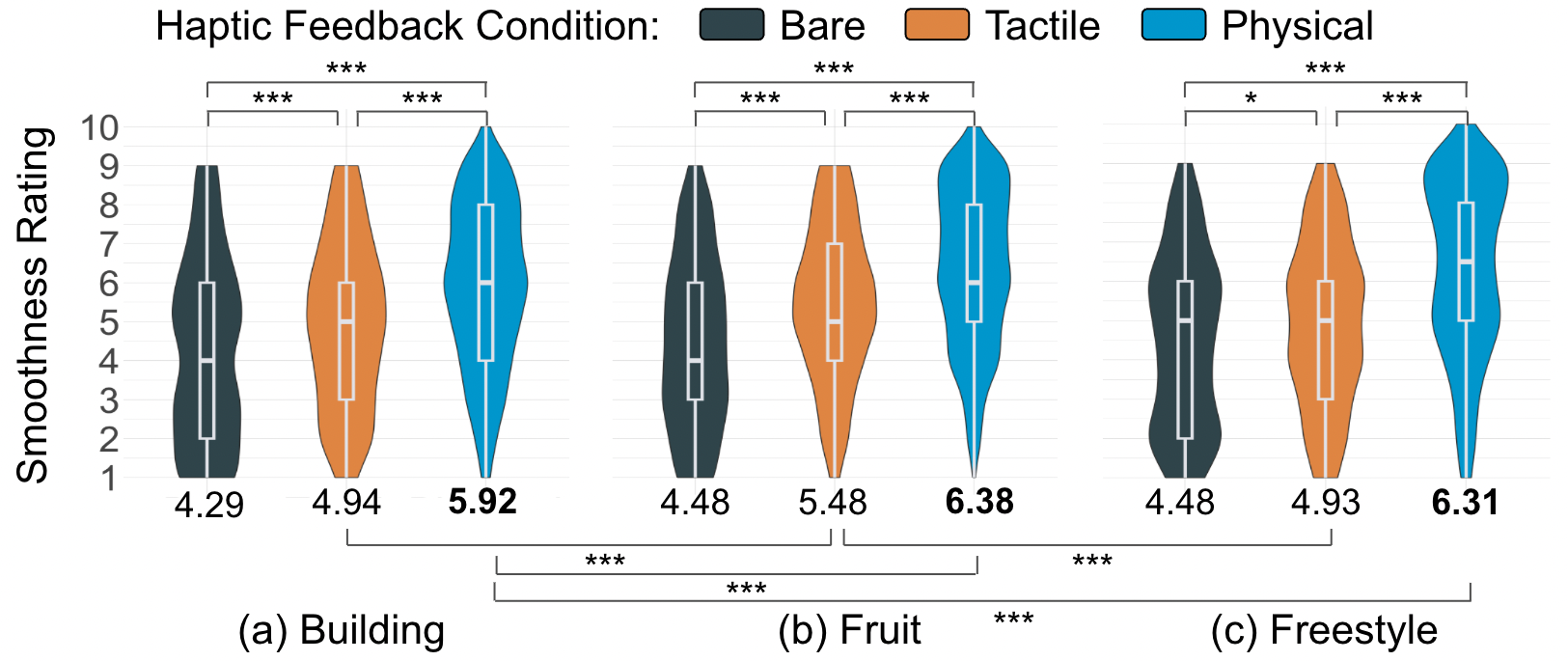}
\caption{Stroke smoothness ratings for sketches drawn under different conditions. Mean values were shown at the bottom of each condition.}
\label{smoothness_rating}
\end{figure}

The haptic condition significantly affected smoothness ratings ($F_{(2,58)}=26.7,\ p<0.001$), with post-hoc analyses revealing significant differences among all haptic conditions $(p < 0.05)$. As shown in Figure \ref{smoothness_rating}, the \textit{Physical} condition yielded the highest smoothness ratings, averaging $5.92$ ($SD=2.08$) for buildings, $6.38$ ($SD=2.00$) for fruits, and $6.31$ ($SD=2.22$) for freestyle sketches. The \textit{Tactile} condition followed, with mean ratings of $4.94$ ($SD=2.14$) for buildings, $5.48$ ($SD=2.00$) for fruits, and $4.93$ ($SD=2.09$) for freestyle sketches. The \textit{Bare} condition resulted in the lowest ratings, averaging $4.29$ ($SD=2.34$) for buildings, $4.48$ ($SD=2.19$) for fruits, and $4.48$ ($SD=2.21$) for freestyle sketches. 

Sketch type also showed a significant effect on stroke smoothness ($F_{(2,58)}=8.2,\ p<0.05$), and a significant interaction was found between sketch type and haptic condition ($F_{(4, 116)}=3.9,\ p<0.05$). Differences between sketch types were significant only under the \textit{Tactile} and \textit{Physical} conditions ($p<0.05$). Participants drew smoother strokes with haptic feedback, particularly for complex sketches like \textit{Fruit} and \textit{Freestyle}, which involve intricate shapes and curves.

\subsubsection{Stroke Continuity}

\begin{figure}[t]
\centering
\includegraphics[width=\linewidth]{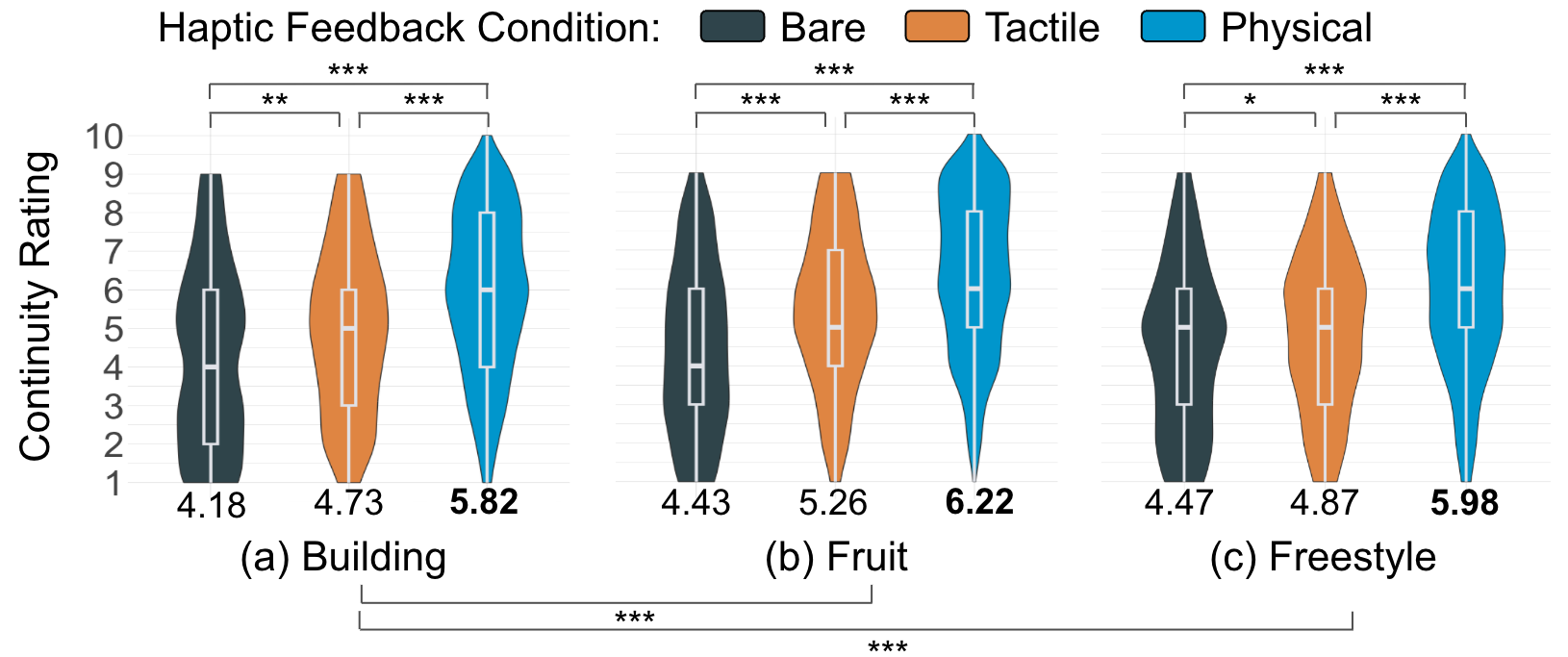}
\caption{Stroke continuity ratings for sketches drawn under different conditions. Mean values were shown at the bottom of each condition.}
\vspace{-8pt}
\label{continuity_rating}
\end{figure}

As illustrated in Figure \ref{continuity_rating}, the haptic condition also had a significant effect on the continuity ratings ($F_{(2,58)}=24.3,\ p<0.001$). The \textit{Physical} condition achieved the highest ratings, averaging $5.82\ (SD=2.02)$ for buildings, $6.22\ (SD=2.08)$ for fruits, and $5.98\ (SD=2.05)$ for freestyle sketches. The \textit{Tactile} condition followed with mean ratings of $4.73\ (SD=2.18)$ for buildings, $5.26\ (SD=1.97)$ for fruits, and $4.87\ (SD=2.02)$ for freestyle sketches. The \textit{Bare} condition received the lowest average ratings at $4.18\ (SD = 2.25)$ for buildings, $4.43\ (SD=2.18)$ for fruits, and $4.47\ (SD=2.13)$ for freestyle sketches. Post-hoc tests showed that the continuity ratings under the three haptic conditions significantly differed ($p<0.05$). 

In addition, sketch type significantly affected stroke continuity ratings ($F_{(2,58)}=8.8,\ p<0.001$). \textit{Fruit} sketches received the highest mean rating of $5.30\ (SD=2.20)$, followed by \textit{Freestyle} sketches at $5.11\ (SD=2.16)$, while \textit{Building} sketches had the lowest mean rating of $4.91\ (SD=2.25)$. Post-hoc tests showed that significant differences occurred only between \textit{Building} sketches and the other two sketch types ($p<0.001$).

\subsubsection{Sketch Clarity}

\begin{figure}[t]
    \centering
    \includegraphics[width=\linewidth]{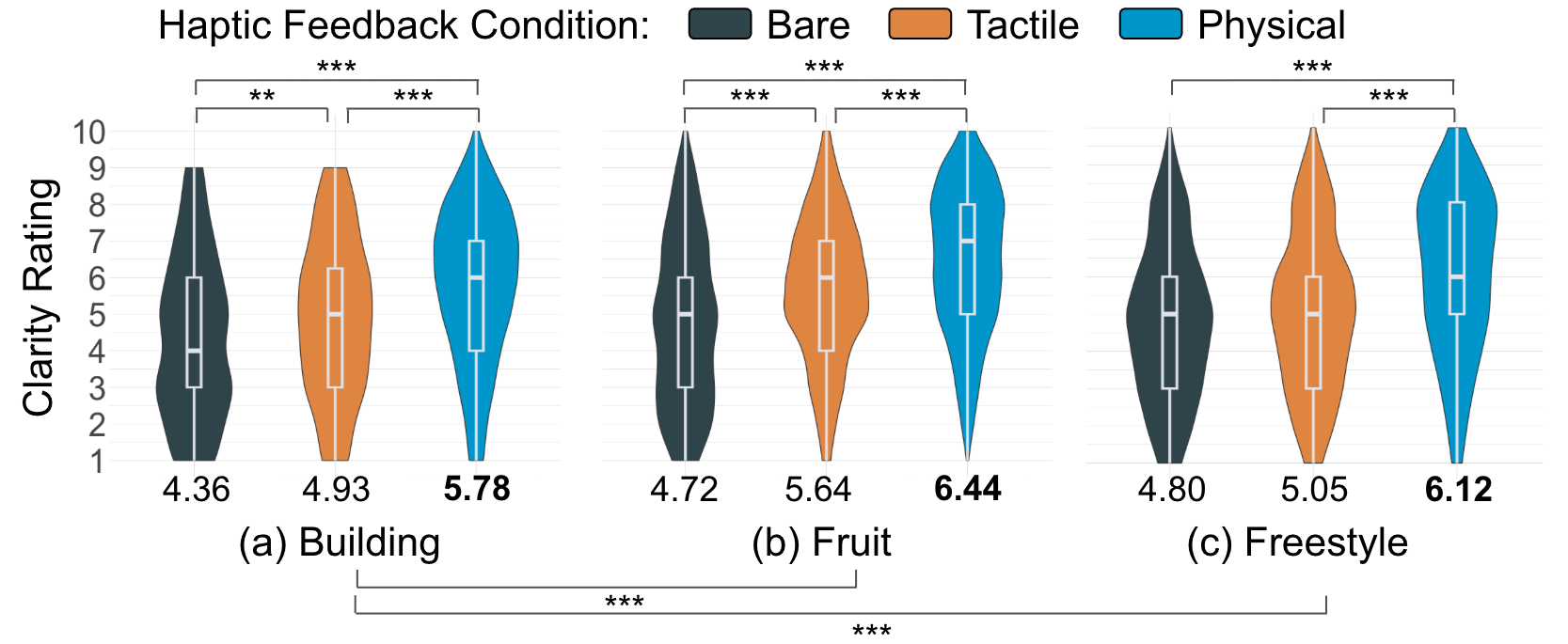}
    \caption{Sketch clarity ratings for sketches drawn under different conditions. Mean values were shown at the bottom of each condition.}
    \label{clarity_rating}
\end{figure}

The haptic condition had a significant effect on sketch clarity ($F_{(2,58)}=21.2,\ p<0.001$). As shown in Figure \ref{clarity_rating}, the \textit{Physical} condition achieved the highest mean ratings: $5.78\ (SD=2.06)$ for buildings, $6.44\ (SD=1.99)$ for fruits, and $6.12\ (SD=2.12)$ for freestyle sketches. The \textit{Tactile} condition followed with mean ratings of $4.93\ (SD=2.16)$ for buildings, $5.64\ (SD=1.94)$ for fruits, and $5.05\ (SD=2.09)$ for freestyle sketches, while the \textit{Bare} condition yielded the lowest ratings at $4.36\ (SD=2.20)$ for buildings, $4.72\ (SD=2.23)$ for fruits, and $4.80\ (SD=2.07)$ for freestyle sketches. Post-hoc tests showed that all these differences among the haptic conditions were statistically significant $(p < 0.001)$.

Clarity ratings also varied significantly by sketch type ($F_{(2,58)}=20.5,\ p<0.001$). \textit{Fruit} sketches had the highest mean clarity rating of $5.60\ (SD=2.17)$, followed by \textit{Freestyle} sketches with $5.32\ (SD=2.19)$, and \textit{Building} sketches with $5.03\ (SD=2.22)$. Post-hoc tests revealed that the \textit{Building} sketches were rated significantly lower than the other two types $(p < 0.001)$.  

\subsubsection{Bimanual Behavior in the Sketching Task}
We also examined participants’ dominant and non-dominant hand movements during sketching to explore coordination patterns. Since the sketching task was less controlled than the selection and tracing tasks, detailed analyses and visualizations are provided in \ref{appendix:sketching_bimanual_behavior}. Under the \textit{Physical} condition, participants tended to begin strokes with their dominant finger farther from the surface and approach it more quickly than in the \textit{Tactile} and \textit{Bare} conditions (Figure \ref{sketch_finger_surface_dis}). The physical surface also led to smaller overall movements of the dominant hand (Figure \ref{sketch_hand_move}) and greater movements of the non-dominant hand (Figure \ref{sketch_surface_move}), suggesting that the physical surface offloaded effort from the dominant hand and engaged the non-dominant hand more actively for coordinated sketching.

\subsection{Participant Feedback}

\subsubsection{Haptic Preference}
Half of the participants preferred the \textit{Physical} condition because it was the same setup people use to interact with ``the real world'', where participants could leverage their existing knowledge of drawing. Interacting with the physical surface helped them complete the VR tasks ``easily, more controllably, and precisely''. Five participants indicated the \textit{Tactile} condition was their favorite, citing its sensitivity and immediate feedback in response to surface contact. However, the noise produced by actuators was reported as being ``distracting''. Some participants complained that interactions under the \textit{Tactile} condition were both physically and mentally demanding because they needed to ``focus on'' the haptic instructions while constantly ``adjusting'' their fingers. Most participants regarded \textit{Bare} as the worst experience, noting uncertainty about ``whether I was doing tasks correctly or not'' and feeling of ``loss of control, patience, and confidence'', despite acknowledging that it offered more ``freedom''.

\subsubsection{Confidence and Fatigue}

\begin{figure}[t]
\centering
\includegraphics[width=\linewidth]{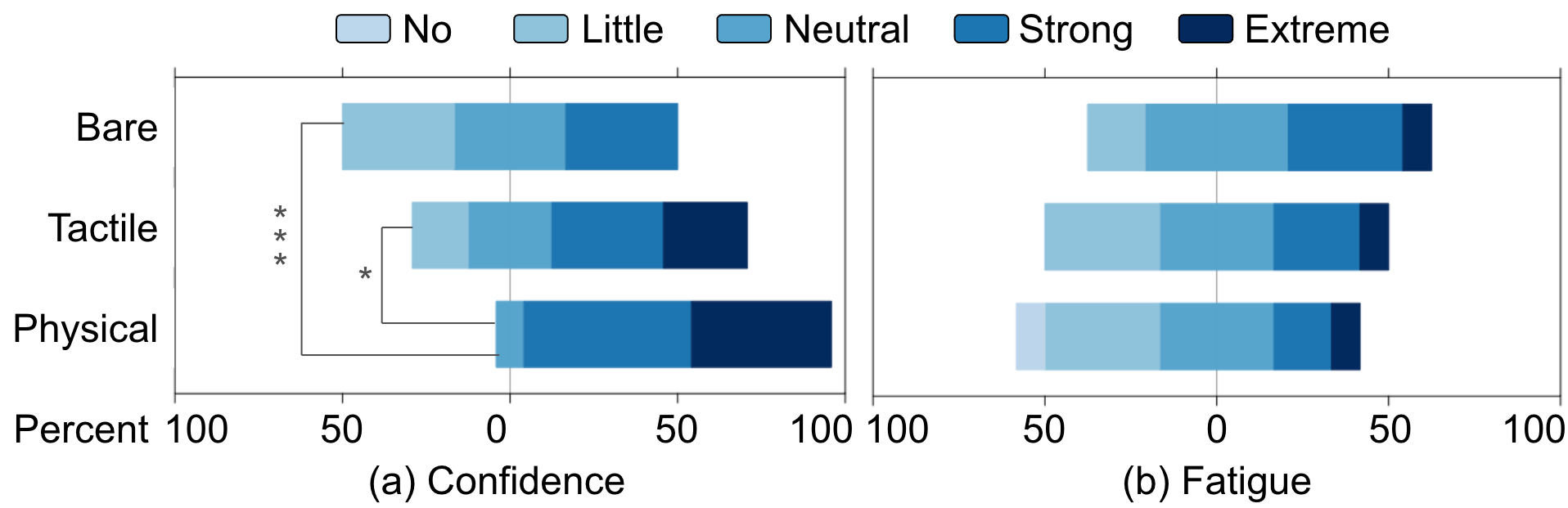}
\caption{ Participants' ratings of: (a) confidence; (b) fatigue when given the three types of haptic feedback.}
\label{ratings}
\end{figure}

Participants reported greater confidence and reduced fatigue when interacting in the \textit{Physical} condition. Confidence differed significantly across haptic conditions ($F_{(2,22)}=9.27,\ p<0.001$), while fatigue did not ($F_{(2,22)}=0.83,\ p>0.05$). This aligns with participants' feedback describing the physical surface as ``natural'', ``intuitive'', and ``easy to control'' (Figure \ref{ratings}).

Regarding fatigue, participants noted that resting the dominant finger on the physical surface helped reduce hand and arm strain. In contrast, with visual-only or tactile feedback, the dominant hand experienced greater fatigue due to increased movement in mid-air. While holding a physical surface or not could influence fatigue, participants did not report that holding the physical plate increased fatigue. Participants identified the weight of the VR headset ($\sim 500$ g) and haptic gloves (316 g per glove) as a major contributor to fatigue, which were more than twice as heavy as the physical plate ($185$ g) and were constant across all conditions, potentially masking differences attributable to haptic feedback.

\subsubsection{Task Load}

\begin{figure}[t]
\centering
\includegraphics[width=\linewidth]{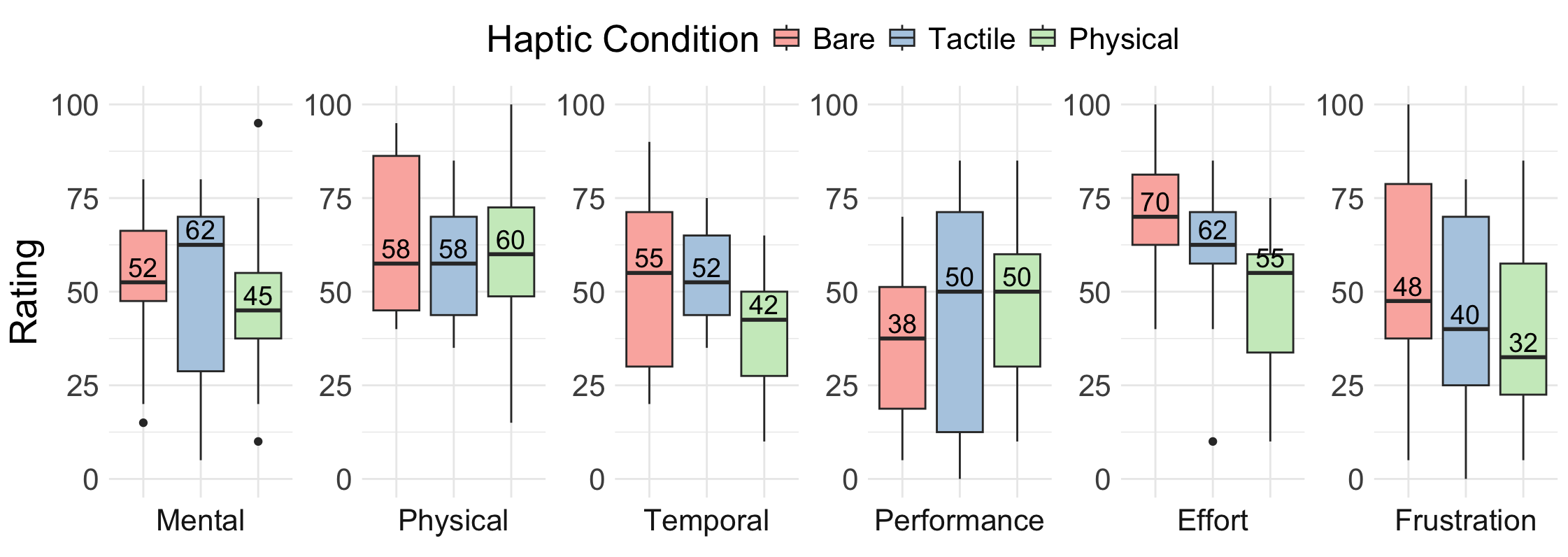}
\caption{Mean NASA TLX ratings for workload factors under the three haptic conditions.}
\vspace{-6pt}
\label{workload}
\end{figure}

The workload ratings for the three types of haptic feedback are presented in Figure \ref{workload}. Although no significant differences were found among the haptic conditions, interactions with the physical surface were perceived as less mentally and temporally demanding, requiring less effort and causing less frustration. In contrast, the \textit{Tactile} condition was rated as more mentally demanding, while the \textit{Bare} condition was considered more frustrating, requiring more effort but resulting in worse performance.

%% file: Discussion.tex
\section{DISCUSSION}
Overall, the \textit{Physical} condition significantly outperformed the \textit{Bare} and \textit{Tactile} conditions in terms of selection accuracy, tracing precision and speed, and sketching quality. The benefits of the physical surface may be more pronounced in complex tasks involving continuous input compared to simpler tasks with discrete input.

Participants exhibited distinct bimanual behaviors across haptic conditions. In the \textit{Physical} condition, they hovered their fingers higher before contact, moved the dominant hand less, and relied more on the non-dominant hand. This suggests that participants' two hands might share the workload more effectively and demonstrate better coordination when interacting with the physical surface. 

Most participants preferred the \textit{Physical} condition due to its ease of control, precise performance, and consistency with real-world interactions. Interacting with a physical surface also helped participants reduce fatigue by allowing them to rest their dominant finger on the surface and reduce dominant hand movement. The \textit{Tactile} condition was ranked second, valued for its responsive tactile and pressure feedback. However, it resulted in a greater mental burden to process the tactile sensations and sometimes required more attention to adjust the dominant hand. The \textit{Bare} condition was reported as the least satisfactory due to performance deficiencies and reduced confidence it introduced. 

\subsection{Effect of Haptic Feedback on Task Performance}
\subsubsection{Selection Task with Discrete Input}
During the selection task, the \textit{Physical} condition resulted in a 50\% reduction in selection errors compared to \textit{Tactile} and \textit{Bare} conditions. Furthermore, the accuracy improvement provided by the physical surface was more pronounced for small-target selection than for large-target selection. The observed differences in selection errors across haptic conditions were significant, diverging from the findings of the HapThimble study \cite{HapThimble}, which reported similar clicking performance across all haptic feedback types. A possible explanation for this discrepancy is that the selection task in our study was more challenging, involving smaller targets on a portable surface, thereby suggesting that physical surfaces may offer greater benefits in high-precision and demanding tasks. 

However, throughput (TP) did not follow the same trend, as the \textit{Physical} condition resulted in slightly lower TP compared to the \textit{Bare} condition. This finding aligns with the HapThimble \cite{HapThimble} study, which reported that selections on a grounded physical surface tended to take longer than those performed in mid-air. This difference could be attributed to mid-air tapping being faster but less accurate, while physical touch provided accuracy and stability at the cost of speed.

\subsubsection{Tracing and Sketching Tasks with Continuous Input}
Tracing on a physical surface achieved the highest speed and levels of stroke precision, smoothness, and continuity. Regarding the sketching task, the \textit{Physical} condition was also rated as the most effective in terms of stroke smoothness, continuity, and clarity. Meanwhile, the \textit{Bare} condition led to the worst task performance in both tracing and sketching tasks, and the \textit{Tactile} condition showed intermediate performance, ranking between the \textit{Bare} and \textit{Physical} conditions. 

In our study, the \textit{Physical} condition showed greater advantages for continuous input tasks (tracing, sketching) than for discrete input tasks (selection), which may be attributed to differences in task complexity. Selection only involves simple up-and-down movements along the surface normal axis (z-axis). In contrast, tracing and sketching maintain touch contact (z-axis) while simultaneously executing continuous movements along the surface's x- and y-axes, incorporating actions such as pointing, dragging, steering, and position control. Our findings suggest that continuous input tasks benefit more from physical support, as users can easily maintain consistent contact with the surface without extra effort to stabilize their hands in mid-air.

Additionally, we found an interesting interaction between haptic conditions and stroke types. While the \textit{Physical} condition provided greater benefits than the \textit{Bare} and \textit{Tactile} conditions for both triangle and circle tracing, its advantages varied depending on whether the lines were straight or curved. Specifically, the physical surface improved precision and continuity in circle tracing, while it enhanced speed in triangle tracing. Similarly, when sketching curves (e.g., fruit shapes), the \textit{Physical} condition provided greater benefits for stroke clarity and smoothness. These findings suggest that using a physical surface is particularly effective for improving the quality of curved lines, enhancing precision, clarity, smoothness, and continuity, while still offering some benefits for straight lines, such as increased speed.

While prior work \cite{NormalTouch_and_TextureTouch, xiao18mrtouch} did not provide evaluations of stroke smoothness, continuity or clarity in tracing or sketching tasks, they showed benefits of using a physical surface for line-based input similar to the findings in our study. For example, using a pen to trace on a grounded surface \cite{evaluation_sketching_Arora} or using a finger with a haptic device \cite{NormalTouch_and_TextureTouch} was found to improve precision. Our tracing precision results also compare favorably with those reported in previous tracing studies. Specifically, MRTouch reported a mean Euclidean distance offset of about 4.0 mm \cite{xiao18mrtouch}, Arora’s work reported a Mean Projected Deviation of 8.4 mm \cite{evaluation_sketching_Arora}, and NormalTouch and TextureTouch showed mean tracing errors of approximately 5–10 mm. However, the experimental setups differed, as our study used a small portable surface, while these studies used large grounded surfaces \cite{evaluation_sketching_Arora, xiao18mrtouch} or fingertip-only haptic devices \cite{NormalTouch_and_TextureTouch}, and the target shapes and sizes also varied. Whether portable physical surfaces can outperform grounded surfaces or haptic devices in tracing or sketching tasks remains an open question for future research.

\subsection{Effect of A Physical Surface on Bimanual Interactions}
In this study, we found that participants engaged in different hand movements under the \textit{Physical} condition compared to \textit{Tactile} and \textit{Bare} conditions. Participants tended to initiate a touch by hovering the manipulation finger of the dominant hand at a greater distance from the surface and approaching the target more quickly. Additionally, participants needed less dominant hand movement to complete a task on the physical surface, while the non-dominant hand showed increased movement. We observed that participants sometimes moved the physical surface slightly toward where they intended to touch. This finding might suggest that the non-dominant hand would likely adjust the surface for better coordination with the dominant hand, and this adjustment could alleviate the workload of the dominant hand, thereby reducing its total dominant hand movement during the task. 

Interestingly, this two-hand coordination pattern was only observed in the \textit{Physical} condition. Based on the user feedback, one possible explanation is that participants felt more comfortable and confident when touching a physical surface, as they noted that they could leverage their existing drawing experience from the real world. Holding a physical surface in hand encouraged them to manipulate the virtual surface more naturally, as it was perceived as easier to control. In contrast, mid-air interactions without physical support were considered less controllable, requiring additional effort to interpret the visual or tactile feedback, address depth perception challenges, and accurately transition between touch and no-touch states. The additional cognitive load may have caused participants to limit surface manipulation with the non-dominant hand to stabilize the virtual surface, thereby increasing movement compensation by the dominant hand to complete the task successfully. 

We hypothesize that interacting with a physical surface in VR can better leverage the benefits of bimanual interactions compared to interacting with visual-only and tactile feedback. A physical surface offers tactile cues, stable resistance, and a kinesthetic frame of reference \cite{10.1145/320719.322599}, enhancing spatial awareness and proprioception \cite{mine1997moving}. Prior work shows that such cues help users gauge hand position and motion relative to the surface \cite{10.1145/320719.322599, mine1997moving, hinckley1998two} fostering improved coordination between hands, more balanced workload, and greater precision with less fatigue \cite{10.1145/191666.191821}.

While using a physical surface offers advantages for bimanual interaction, it may also constrain the movement of non-dominant hand, which is occupied with holding the surface. Nevertheless, even while gripping the plate, the non-dominant hand can still perform simple thumb interactions \cite{wagner2012bitouch, pfeuffer2017thumb+} and microgestures \cite{sharma2021solofinger}, for input confirmation, shortcuts, and changing input modes. This balance between stability and flexibility underscores the broader potential of physical surfaces in supporting precise yet versatile bimanual interaction in VR.

\subsection{Enabling Physical Surfaces for Bimanual Interaction in VR}

Our study highlighted the advantages of incorporating physical surfaces in VR, particularly in enhancing interaction performance and improving coordination between the two hands. Although carrying or holding a surface can introduce some setup effort, the benefits it affords for precise and coordinated manipulation may outweigh the inconvenience, particularly in productivity and creative tasks that demand accuracy and endurance. Scenarios such as VR sketching, annotating, or on-window interaction could all benefit from the fine control and familiar experience that physical surfaces provide. 

Beyond performance benefits, long-term use of physical surfaces may enhance proprioceptive consistency and embodied skill development, enabling users to perform interactions more efficiently over time \cite{mine1997moving}. Repeatedly interacting with a physical surface allows users to build muscle memory and engage in more eye-free control, similar to how skilled tablet or keyboard users rely on spatial familiarity \cite{brewster2003multimodal, yi2012exploring, zhu2019isfree}. These possibilities highlight the potential of a simple physical surface to enhance expert-level fluency in complex VR tasks.

Besides the simple hand-held physical surface used in this study, previous research has shown that everyday objects with physical surfaces, ranging from touchscreen-equipped devices \cite{evaluation_sketching_Arora, SymbiosisSketch, VRSketchIn, TabletInVR} to everyday objects \cite{10.1145/332040.332494, xiao18mrtouch}, can effectively serve as haptic proxies for precise interaction in VR. Looking ahead, advances in hand tracking and spatial sensing could enable seamless touch detection on arbitrary real surfaces, allowing everyday items such as notebooks, folders, or tabletops to become interactive without additional instrumentation. Such developments would make physical surfaces an integral component of productivity-oriented and creative VR experiences, combining realism, precision, and long-term ease of use. 

\subsection{Limitations and Future Work}

\subsubsection{Effects of Surface and Hand Characteristics}

While we utilized a physical surface in this study, such surface consists of multiple attributes, including traction, texture, support, and synchronized feedback across hands, each of which may contribute differently to manipulation performance. Identifying these factors requires further investigation. 

This study focused on single-point, two-handed touch input on a flat surface, but future work could explore multi-touch, one-handed input, or interactions on non-planar surfaces. We examined movement speed and distance to assess how physical surfaces influence bimanual coordination. While informative, more detailed analyses, such as 3D trajectories of the hands, fingers, and surface, could provide deeper insight into bimanual interaction. Additionally, we did not collect detailed hand anthropometric data, such as hand size or finger length. Future work could examine how individual hand characteristics affect interaction performance and user experience.

While this study examined portable surface interactions with participants seated throughout the experiment, future work could investigate how different postures, such as standing, sitting, or mid-motion, affect user performance and experience under varying haptic conditions to better understand the role of mobility in real-world VR applications.

\subsubsection{Device and Ergonomic Constraints}
Equipment-related constraints such as device weight and tracking accuracy, and gesture ergonomics, can influence fatigue and overall experience. The headset and haptic gloves added load to the user's body which sometimes caused discomfort during prolonged use or tasks involving sustained mid-air gestures. 
While rarely observed in the study, tracking issues, such as drift or latency, could also influence performance and comfort because they necessitate corrective finger movements. 
These factors represent limitations of the current setup. Future systems with lighter hardware, more accurate and reliable tracking, and improved ergonomics may reduce fatigue and improve the usability of physical-surface interactions in VR.

\subsubsection{Participant Expertise and Learning Effects}
The participants in the study were novices to VR sketching and had no professional sketching experience, which may limit the generalizability of our findings to expert artistic practice. However, sketching and annotation are common tasks for everyday users, and we believe that it is important to understand how physical surfaces support novice performance. For freestyle sketches, we recruited independent reviewers to assess sketch quality using basic stroke-related measures (smoothness, continuity, clarity) following prior studies \cite{evaluation_sketching_Arora, wiese2010invest_free_hand_sketching}, rather than focusing on the artistic quality. This approach reflects how non-expert users perceive sketch quality in practical scenarios. Involving professional sketchers and adopting expert-based evaluation criteria could further extend the assessment to professional sketching, artistic expression, and workflow efficiency in expert creative contexts. Future work could also compare novices and experts to examine learning effects across haptic modalities, vary training duration, and model learning curves to evaluate long-term performance.

\subsubsection{Exploration of Other Interaction Modalities}
Although this study examined how physical surfaces support hand-based interaction in VR, other input tools such as styluses can also be used \cite{SymbiosisSketch, VRSketchIn}, each offering distinct advantages for different tasks. Pen-based input is widely used for precision activities such as drawing or modeling due to its fine control \cite{finger_pen_stroke, 10.1145/2797138}, but it also introduces practical constraints related to device setup. In contrast, direct hand interaction is immediately accessible, requires no additional hardware, and allows quick, spontaneous engagement \cite{finger_pen_stroke, 10.1145/2797138}. Our results show that, when supported by physical surfaces, hand-based input can achieve high performance and efficiency even in fine-grained tasks, indicating that the hand alone can serve as an effective VR input modality. Physical surfaces can also bridge these two approaches by serving as a shared substrate that supports casual, always-available hand input while accommodating stylus use for higher precision \cite{10.1145/1866029.1866036}. Future work could explore hybrid systems that allow seamless switching between direct hand and tool-mediated interaction based on task demands.

%% file: Conclusion.tex
\section{CONCLUSION}
This paper investigated the effect of different types of haptic feedback on precise finger-based touch input. The findings revealed that incorporating physical interaction surfaces significantly enhanced task performance across selecting, tracing, and sketching activities. Specifically, compared to tactile feedback, the physical surface resulted in a 51.2\% improvement in selection accuracy, a 20.3\% increase in tracing precision, an 86.8\% rise in tracing continuity, and an 18.3\% increase in tracing speed. For the sketching task, the physical surface improved by 21.6\% stroke smoothness, 21.2\% stroke continuity, and 17.5\% sketch clarity. 
Additionally, manipulating a physical surface enables users to coordinate both the surface and their hands more effectively compared to other conditions, facilitating bimanual interactions. 
We hope these insights will inform the design of future haptic interfaces, encouraging the integration of physical surfaces to support precise and intuitive hand interactions in VR environments.

%% file: Appendix.tex
\section{Sketching Rating Criteria}
\label{appendix:sketching_rating_criteria}
\begin{figure}[t]
\centering
\includegraphics[width=\linewidth]{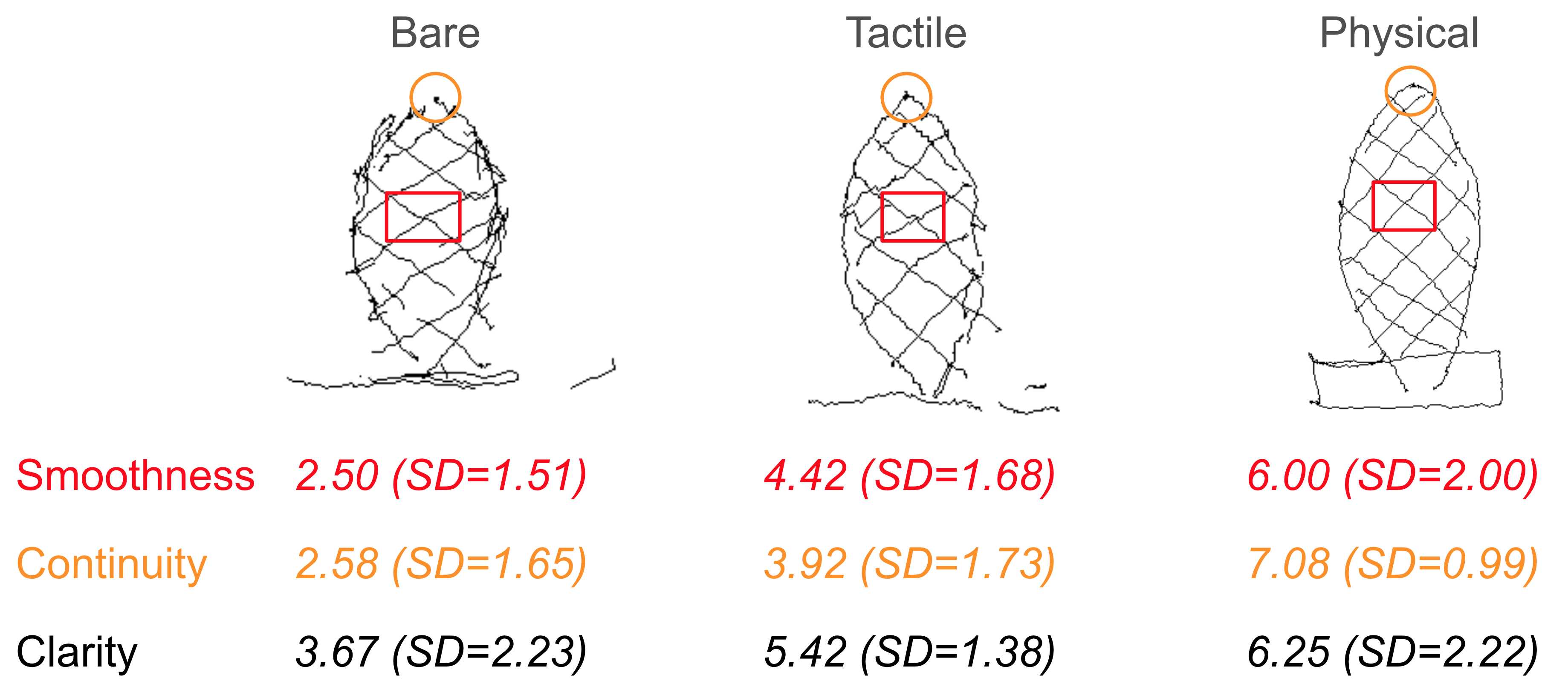}
\caption{Mean ratings of stroke smoothness, continuity, and clarity for textured-building sketches under different conditions. We marked areas that may indicate areas demonstrating different stroke smoothness and stroke continuity in red boxes and orange circles, respectively. 
}
\label{building_rating_exp}
\end{figure}

\begin{figure}[t]
\centering
\includegraphics[width=\linewidth]{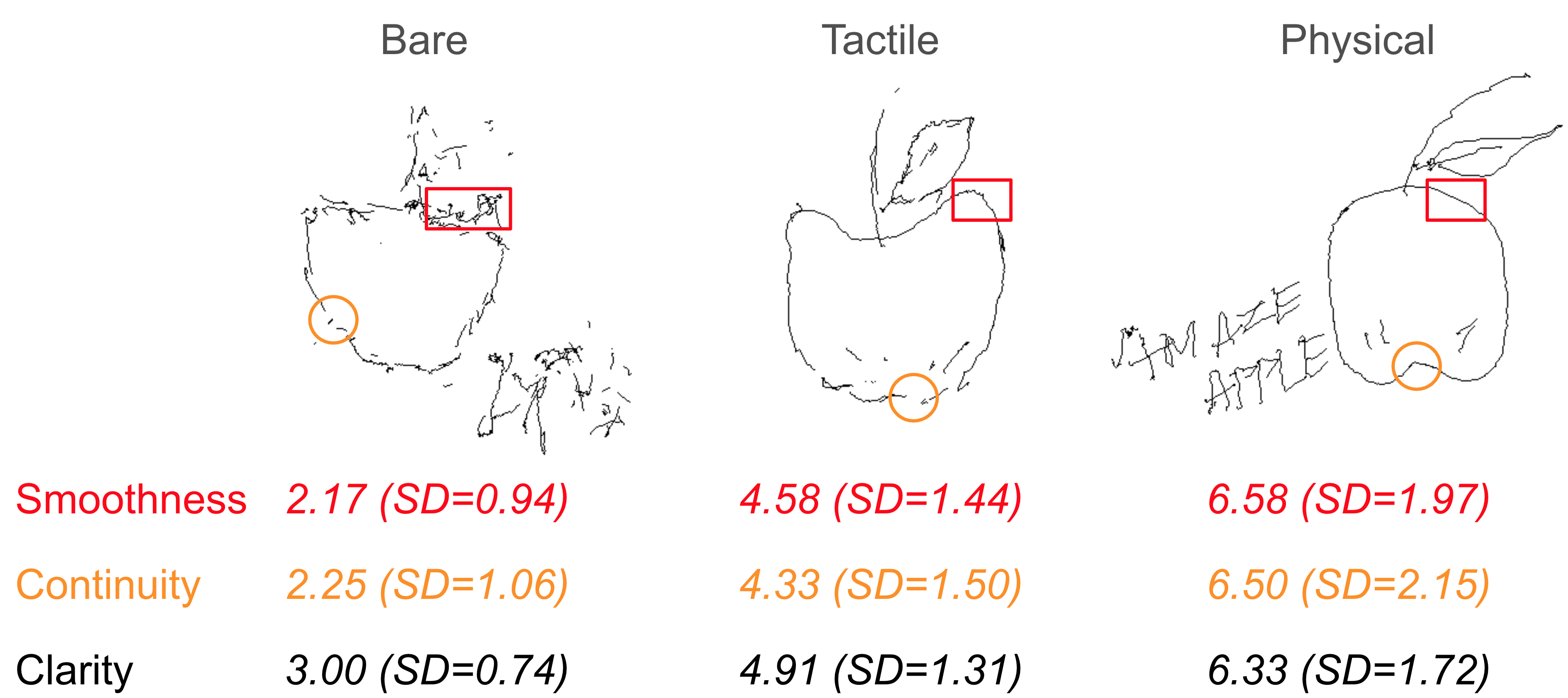}
\caption{Mean ratings of stroke smoothness, continuity, and clarity for fruit-shape sketches under different conditions. We marked areas that may indicate areas demonstrating different stroke smoothness and stroke continuity in red boxes and orange circles, respectively.}
\label{fruit_rating_exp}
\end{figure}

\begin{figure}[t]
\centering
\includegraphics[width=\linewidth]{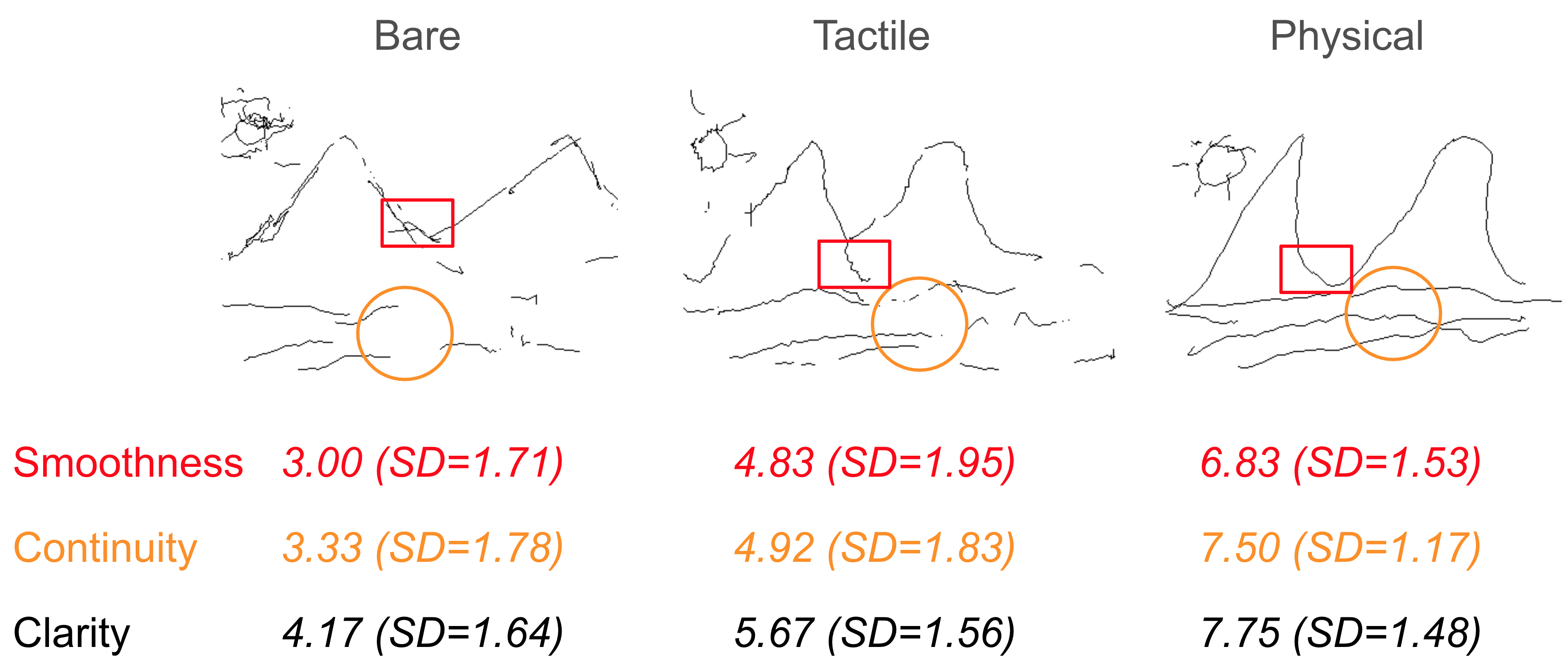}
\caption{Mean ratings of stroke smoothness, continuity, and clarity for freestyle sketches under different conditions. We marked areas that may indicate areas demonstrating different stroke smoothness and stroke continuity in red boxes and orange circles, respectively.}
\label{freestyle_rating_exp}
\end{figure}

Examples from three participants in the (1) textured-building (Figure \ref{building_rating_exp}), (2) fruit-shape (Figure \ref{fruit_rating_exp}), and (3) freestyle (Figure \ref{freestyle_rating_exp}) sketching tasks, along with external reviewers’ ratings of stroke smoothness, stroke continuity, and overall clarity. Portions of the sketches illustrating smoothness and continuity are annotated for reference.

\section{Bimanual Behavior in the Sketching Task}
\label{appendix:sketching_bimanual_behavior}

\subsection{Dominant Hand Trajectory}

\begin{figure}[t]
\centering
\includegraphics[width=\linewidth]{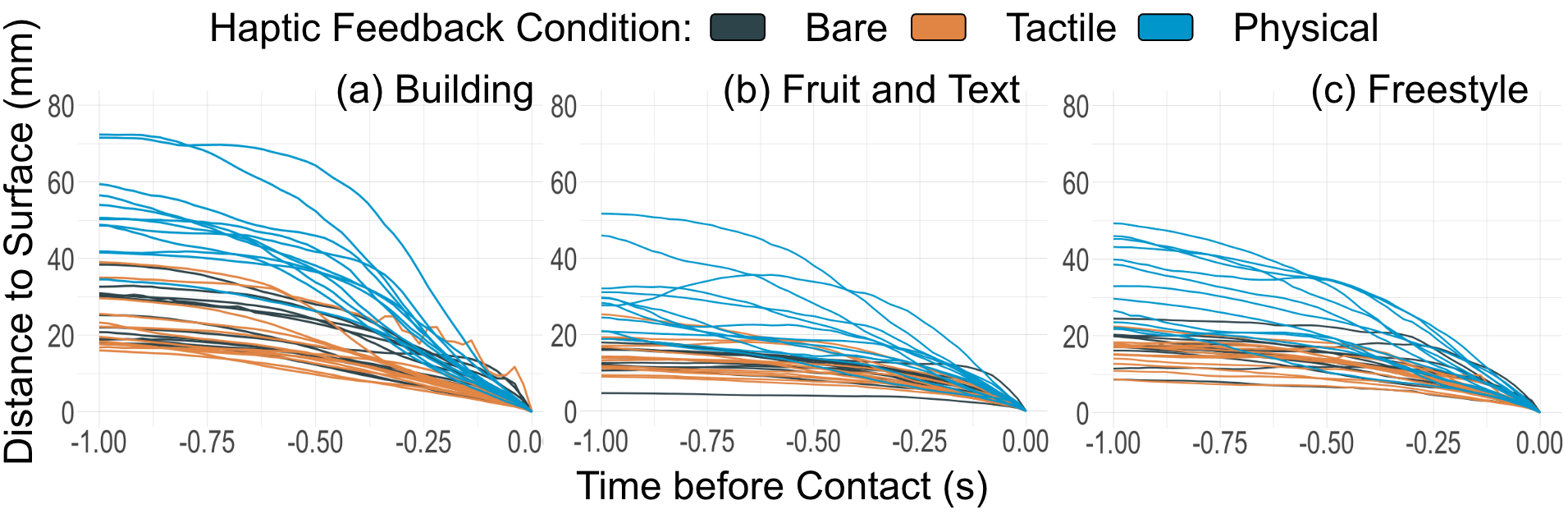}
\caption{ The dominant hand trajectory along the surface’s normal direction (i.e.,
finger-surface distance) within one second before contact. The sketching task used three configurations: (a) a textured building, (b) a fruit with text, and (c) a freestyle sketch.}
\vspace{6pt}
\label{sketch_finger_surface_dis}
\end{figure}

The dominant hand trajectory along the surface's normal direction (i.e., finger-surface distance) before contact was illustrated in Figure \ref{sketch_finger_surface_dis}. Both the haptic condition ($F_{(2,22)}=47.5,\ p<0.001$) and the target size ($F_{(2,22)}=21.0,\ p<0.001$) showed a significant effect on this trajectory, and an interaction was also found between them ($F_{(4,44)}=7.69,\ p<0.001$). 

The most significant difference in finger-surface distance was observed at $0.75\ s$ before contact during all types of sketching. Specifically, participants held the dominant finger farthest to the surface to start a stroke when sketching a textured building ($M=70.0\ mm,\ SD=34.0\ mm$), followed by freestyle sketching ($M=48.8\ mm,\ SD=21.3\ mm$) and fruit \& text-sketching ($M=42.5\ mm,\ SD=19.8\ mm$). Post-hoc tests revealed the differences among these three types of sketches were all significant ($p<0.05$).

Regarding haptic conditions, under the \textit{Physical} condition, most participants initiated a stroke by holding their dominant finger significantly farther away from the surface, then approaching the surface with a higher speed compared to \textit{Tactile} and \textit{Bare} conditions ($p<0.001$). 

Under the \textit{Physical} condition, participants held their dominant finger at a mean distance of $49.1\ (SD=10.8)\ mm$ from the surface during building-sketching, $26.7\ (SD=9.88)\ mm$ during fruit-sketching and $29.5\ (SD=10.4)\ mm$ during the freestyle sketching. This finger-surface distance was approximately $100\%$ larger than the distance observed under the \textit{Tactile} and \textit{Bare} conditions. The \textit{Tactile} condition resulted in a mean finger-surface distance of $21.3\ (SD=7.51)\ mm$ for building-sketching, $13.4\ (SD=4.27)\ mm$ for fruit-sketching and $14.5\ (SD=3.73)\ mm$ for the freestyle sketching. Similarly, the \textit{Bare} condition led to a mean distance of $23.7\ (SD=6.52)\ mm$ for building-sketching, $13.2\ (SD=3.57)\ mm$ for fruit-sketching and $17.7\ (SD=5.98)\ mm$ for the freestyle sketching. Post-hoc tests revealed the differences among these three haptic conditions were all significant ($p<0.05$).

\subsection{Total Dominant Hand Movement}

\begin{figure}[t]
\centering
\includegraphics[width=\linewidth]{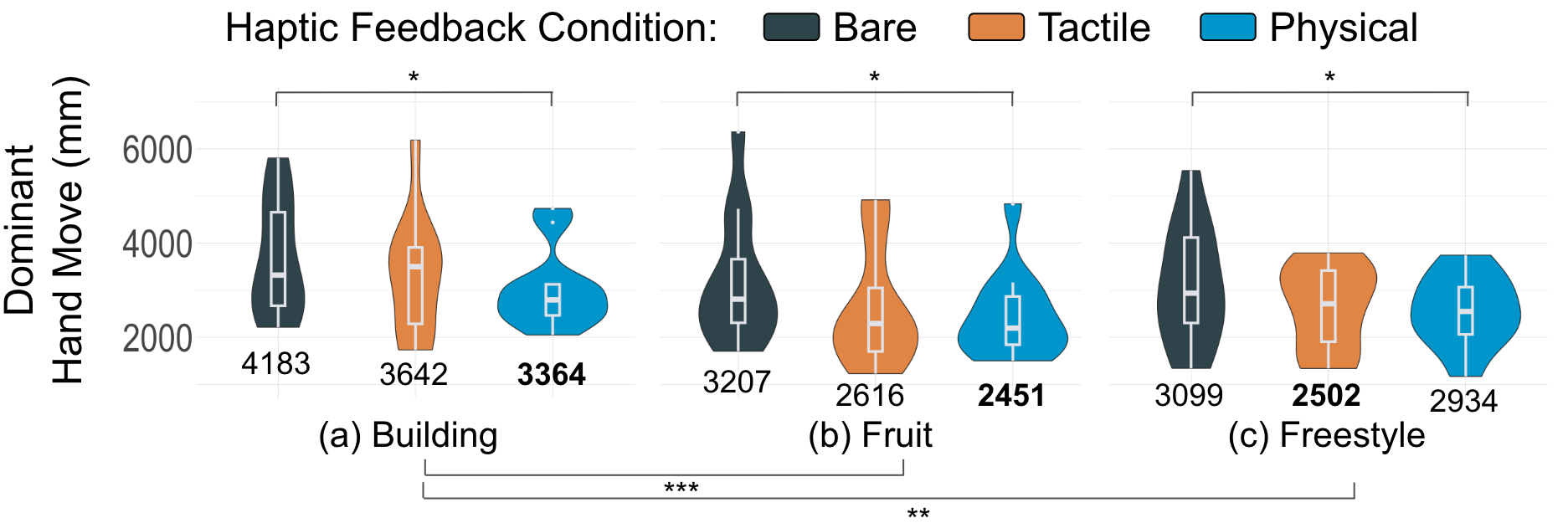}
\caption{ The total dominant hand movement when sketching one (a) textured building, (b) fruit shape with some text, and (c) freestyle sketch under different conditions. Mean values were shown at the bottom of each haptic condition.}
\vspace{4pt}
\label{sketch_hand_move}
\end{figure}

As shown in Figure \ref{sketch_hand_move}, the haptic condition had a significant impact on the total dominant hand movement for completing a sketch $(F_{(2,22)}=4.02,\ p<0.05)$. With the physical surface, participants moved the dominant finger $3364.3\ (SD=1412.0)\ mm$ when sketching a textured building, $2451.2\ (SD=928.2)\ mm$ when sketching a fruit shape with text, and $2934.2\ (SD=2021.7)\ mm$ for freestyle sketching. The \textit{Tactile} condition followed, where participants moved the dominant finger an average distance of $3642.5\ (SD=1684.4)\ mm$ for building-sketching, $2616.1\ (SD=1261.4)\ mm$ for fruit-sketching, and $2502.1\ (SD=974.9)\ mm$ for freestyle sketching. The largest dominant hand movement was observed under the \textit{Bare} condition, with participants drawing a building, fruit with text, and a freestyle sketch over an average distance of $4182.7\ (SD=2202.2)\ mm$, $3207.4\ (SD=1370.7)\ mm$, and $3099.2\ (SD=1255.9)\ mm$, respectively. However, post-hoc tests found only the \textit{Physical} and \textit{Bare} conditions were significantly different from each other ($p<0.05$).

In addition, the type of sketch significantly influenced the amount of dominant hand movement during sketching $(F_{(2,22)}=9.44,\ p<0.001)$. Post-hoc tests found that building sketching resulted in significantly more dominant hand movement compared to other types of sketches ($p<0.01$). In contrast, no significant difference was found between fruit sketching and freestyle sketching ($p>0.05$). Specifically, sketching a textured building resulted in the most dominant hand movement with a mean distance of $3729.8\ (SD=1778.0)\ mm$, while sketching a fruit with text had a mean value of $2758.3\ (SD=1212.3)\ mm$ and freestyle sketching had a mean value of $2845.2\ (SD=1464.3)\ mm$. This difference is predictable since these three types of sketches involved different amounts and lengths of strokes. For instance, the building sketch needed more strokes to present the texture, contributing to more dominant finger movement.

\subsection{Total Non-Dominant Hand Movement}

\begin{figure}[t]
\centering
\includegraphics[width=\linewidth]{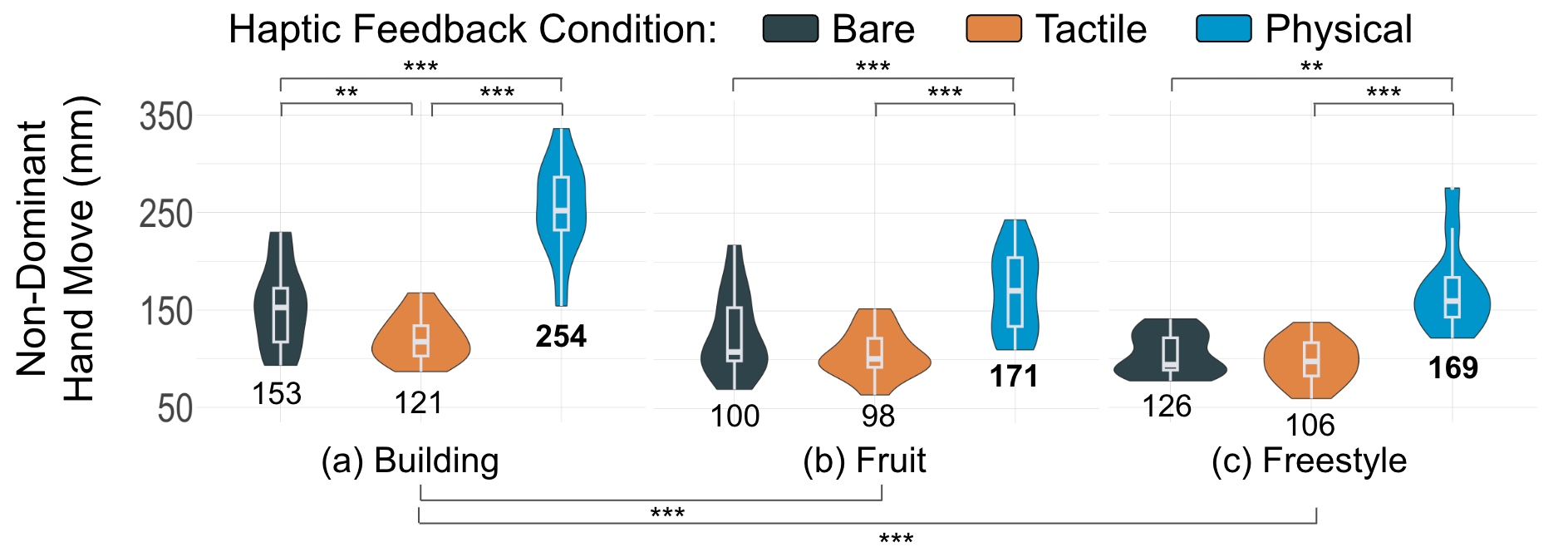}
\caption{ The total non-dominant hand movement when sketching one (a) textured building, (b) fruit shape with some text, and (c) freestyle sketch under different conditions. Mean values were shown at the bottom of each haptic condition.}
\vspace{4pt}
\label{sketch_surface_move}
\end{figure}

Haptic condition significantly influenced total non-dominant hand movement (i.e., surface movement) during sketching $(F_{(2,22)}=37.4,\ p<0.001)$. Sketching on the physical surface resulted in the greatest non-dominant hand movement, as shown in Figure \ref{sketch_surface_move}. The average total non-dominant hand movement distance under the \textit{Physical} condition was $254.3\ (SD=49.3)\ mm$ for building-sketching, $171.5\ (SD=44.6)\ mm$ for fruit-sketching, and $168.8\ (SD=43.9)\ mm$ for freestyle sketching. In contrast, under the \textit{Bare} condition, the mean non-dominant hand movement distance was $153.4\ (SD=44.3)\ mm$ for building-sketching, $99.8\ (SD=27.9)\ mm$ for fruit-sketching, and $126.1\ (SD=43.1)\ mm$ for freestyle sketching. The least non-dominant hand movement was observed under the \textit{Tactile} condition, with participants moving the non-dominant hand $120.8\ (SD=23.8)\ mm$ when sketching a building, $97.9\ (SD=23.0)\ mm$ when sketching a fruit with text, and $106.1\ (SD=24.7)\ mm$ when drawing a freeform sketch. The differences among all haptic conditions were statistically significant $(p<0.01)$.

Significant effects were also found in sketch type for non-dominant hand movement $(F_{(2,22)}=24.2,\ p<0.001)$, and it had an interaction with the haptic condition $(F_{(4,44)}=6.9,\ p<0.01)$. Post-hoc tests found all differences were significant $(p<0.001)$ except for the one between the fruit-sketching and freestyle sketching $(p>0.05)$. In general, sketching a textured building involved most non-dominant hand movement ($M=176.2\ mm,\ SD=69.9\ mm$), followed by freestyle sketching ($M=133.6\ mm,\ SD=45.6\ mm$) and fruit sketching ($M=123.1\ mm,\ SD=47.4\ mm$).